\documentclass{bio}

\usepackage{xcolor}
\usepackage{array}
\usepackage{float}
\newcommand{\belowtablecaption}{\par\setlength{\abovecaptionskip}{5pt}\caption}
\usepackage{url}
\usepackage[margin=1in,headheight=8pt]{geometry}
\makeatletter
\renewcommand\section{\@startsection{section}{1}{\z@}%
  {1.25pc}   
  {6pt}     
  {\normalfont\normalsize\scshape\centering}}

\renewcommand\subsection{\@startsection{subsection}{2}{\z@}%
  {1.25pc}
  {4pt}
  {\normalfont\itshape\centering\noindent}}
\makeatother

\begin{document}

\title{A Bayesian Framework for Latent Compliance Modeling in Cluster Randomized Trials with One-Sided Noncompliance}

\author{ANTHONY SISTI$^{1,\ast}$, ELLEN MCCREEDY$^2$, ROEE GUTMAN$^3$\\[4pt]
\textit{$^1$Global Biostatistical Science, Amgen, Thousand Oaks, CA}\\
\textit{$^2$Department of Health Services, Policy \& Practice, Brown University School of Public Health, 121 South Main Street, Providence, RI 02903}\\
\textit{$^3$Department of Biostatistics, Brown University School of Public Health, 121 South Main Street, Providence, RI 02903}
\\[2pt]
{anthony\_sisti@brown.edu}}

\markboth%
{A. Sisti, E. McCreedy and R. Gutman}
{Bayesian Latent Compliance Model for Cluster RCTs with One-Sided Noncompliance}

\maketitle

\footnotetext{To whom correspondence should be addressed.}

\begin{abstract}
{In pragmatic cluster randomized controlled trials (PCRCTs), healthcare providers are randomized, but intervention delivery and patient receipt may vary after assignment. In many PCRCTs, cluster-level implementation is measured using continuous implementation metrics, while individual compliance is recorded as a binary indicator. Standard complier average causal effect (CACE) estimands focus on individual-level compliance and do not account for heterogeneity in implementation across clusters. When intervention uptake is shaped by both cluster- and individual-level processes, it is of scientific interest to characterize how effects vary across these sources of compliance. We propose a Bayesian framework for PCRCTs with one-sided binary noncompliance at the individual level and one-sided partial compliance at the cluster level. The method uses a latent mixture model to summarize heterogeneity in cluster-level implementation based on baseline characteristics and treatment-arm implementation measures, and links these latent implementation strata to individual compliance and outcomes through a joint model. Because individual compliance and treatment-defined implementation measures are observed only under treatment assignment, the model imputes these quantities for control clusters and individuals assigned to control. The framework enables estimation of finite- and super-population intent-to-treat (ITT) and CACE estimands, both marginally and within latent implementation strata. We apply the method to the METRIcAL trial, a pragmatic cluster randomized study evaluating a personalized music intervention for nursing home residents with dementia. The analysis illustrates how accounting for implementation heterogeneity and individual compliance can provide insights beyond standard ITT analyses.}{Causal inference; Principal stratification; Complier average causal effect; Cluster randomized trials; Noncompliance; Bayesian methods; Latent variable models; Interference.}
\end{abstract}

\section{Introduction}

Agitation and aggressive behaviors are common among cognitively impaired individuals residing in nursing homes (NHs) \citep{Zuidema2007}. These behaviors can decrease quality of life and accelerate cognitive decline \citep{Garcia2019, Shin2005}, while also serving as a major source of stress for NH staff \citep{Karlsen2023} and other residents \citep{Rosen2008}. Antipsychotic and mood-stabilizing medications are commonly used to manage aggressive behaviors; however, these medications are associated with adverse events, including falls and increased mortality \citep{Huybrechts2012}. Thus, there is interest in evaluating non-pharmaceutical interventions that mitigate aggressive behaviors and reduce the side effects of antipsychotic medications.

METRIcAL is a multi-stage, pragmatic, NH-based cluster randomized controlled trial (PCRCT) evaluating a personalized music intervention among NH residents with moderate to severe dementia \citep{Mor2021}. As in many pragmatic trials involving individuals with dementia, a substantial proportion of participants assigned to the intervention did not comply with the intervention protocol \citep{Allore2020}.

Pragmatic randomized controlled trials (PRCTs) are designed to estimate intervention effects under real-world conditions, where participants and providers may vary in their compliance \citep{Patsopoulos2011}. Analyses of PRCTs typically follow the intention-to-treat (ITT) principle, analyzing units according to assigned treatment \citep{Scheiner1995, Gupta2011}. Although ITT estimands preserve randomization, noncompliance can complicate their interpretation and reduce efficiency, particularly in cluster randomized trials where compliance behavior may itself be clustered \citep{Jo2008ITT}. ITT estimands may not characterize effects of intervention receipt or treatment-effect heterogeneity across settings with different compliance patterns \citep{Hernan2017}. Naive alternatives such as as-treated and per-protocol analyses condition on observed receipt and can be biased because they break the original randomization \citep{Ellenberg1996, Matilde2006, Goldberg2014}.

With binary compliance, the complier average causal effect (CACE) preserves the randomization structure by describing the intervention effect among units that would comply with their assigned treatment \citep{Imbens1997}. Because compliance is observed only under assigned treatment, CACE estimation requires additional identifying assumptions. Under monotonicity, exclusion restrictions, and individual-level randomization, instrumental variable (IV) methods can estimate the CACE \citep{Angrist1996}. Bayesian approaches have shown favorable operating characteristics relative to classical IV estimators and provide a framework for sensitivity analyses \citep{Imbens1997, Hirano2000}. Multiple imputation (MI) approaches have also been used to impute missing compliance status, with improved small-sample performance in some settings \citep{Taylor2009}. For cluster randomized designs, existing methods estimate CACEs under individual-level noncompliance \citep{Forastiere2016, Imai2021}, including Bayesian extensions \citep{FrangakisRubinZhou2002, Jo2008CRT, Taylor2011, Ohnishi2022}. However, in pragmatic cluster randomized trials where healthcare providers are randomized, noncompliance may arise at both the cluster and individual levels.

The principal stratification framework provides a flexible approach for defining causal estimands that account for post-treatment compliance \citep{FrangakisRubin2002}. With binary compliance at both the cluster and individual levels, \citet{SchrochetChiang2011} defined a set of principal strata corresponding to all possible compliance combinations and derived moment-based estimators under specific identification assumptions. Subsequent work extended these ideas using likelihood-based approaches under alternative identifying restrictions \citep{Sheng2019}. In many pragmatic trials, however, compliance may be partial rather than binary. For example, individuals may receive different doses of an intervention \citep{Jin2008}, and clusters may vary in the intensity or scope of implementation. To address this \citet{Zhai2020} proposed a Bayesian MI approach that imputes partial cluster-level compliance and binary individual-level compliance, but relied on a single discretized compliance measure and did not specify a joint model for compliance and outcomes.

We propose a Bayesian framework for analyzing PCRCTs with one-sided binary noncompliance at the individual level and one-sided partial compliance at the cluster level. The method uses a latent mixture model to characterize heterogeneity in cluster-level implementation based on pre-treatment characteristics and observed treatment-arm implementation measures. We refer to the mixture components as latent implementation strata. The strata are linked to individual-level compliance and outcomes through a joint model, allowing intervention effects to be estimated marginally and within implementation strata. Because individual-level compliance and treatment-defined implementation measures are observed only under treatment assignment, the model imputes these quantities for clusters and individuals assigned to control. The proposed approach enables estimation of finite- and super-population ITT and CACE estimands.

We apply the method to METRIcAL to examine the effects of a personalized music intervention on agitated behaviors and antipsychotic use among NH residents with dementia. While the 95\% credible intervals for the estimated effects include 0, the analysis illustrates how accounting for cluster-level implementation heterogeneity and individual compliance can yield insights that are not apparent from ITT analyses alone.

\section{Motivating Example: METRIcAL}
\label{METRIcALDescrip}

METRIcAL was a PCRCT of a personalized music intervention for agitated behaviors and antipsychotic use among nursing home (NH) residents with dementia. Stage~1 of METRIcAL included 976 residents across 54 NH facilities, with an average of 18.1 (SD = 1.5) residents per facility. Half of the facilities were assigned to the intervention and half to standard care \citep{Mor2021}. Participants in intervention facilities received iPods containing music they had listened to as young adults, and standard care may include ambient or group music.

In the first stage of METRIcAL,  statistically significant effects  of the personalized music intervention on agitated behaviors were not observed \citep{McCreedy2022}. Limited adherence to the assigned intervention may partly explain this result. Only 71\% of individuals in the intervention arm listened to personalized music at least once, and the median listening time was 22.1 minutes per day, below the prescribed 30 minutes per day \citep{McCreedy2022}.

\subsection{Description of Data}

Facility-level baseline characteristics were obtained from LTCfocus, a long-term care database sponsored by the National Institute on Aging (1P01AG027296) through collaboration with Brown University.
These characteristics included the previous year’s average daily census, racial and gender demographics, geographic location within the United States, average resident age, and the proportion of self-ambulating residents.

Participant-level baseline characteristics were obtained from the Minimum Data Set (MDS) \citep{MDS2021}.
The MDS contains information on residents' health status and functional capabilities, including demographics, comorbid conditions, functional status measured by the Activities of Daily Living Scale (ADL) \citep{Morris1999ADL}, cognitive status measured by the Cognitive Function Scale (CFS) \citep{Thomas2017}, and dementia-related symptoms measured by the Agitated and Reactive Behavior Scale (ARBS) \citep{McCreedy2019}. Appendix Table~\ref{tab:XSummary} summarizes baseline characteristics of the METRIcAL cohort.

Study outcome measures included the Cohen--Mansfield Agitation Inventory (CMAI) and antipsychotic use, which were collected at baseline and four months after study initiation for each participant.
The CMAI is a standardized questionnaire that assesses agitation in older adults and measures 29 agitated behaviors categorized as physically aggressive, physically non-aggressive, or verbally agitated.
Each behavior is rated on a 7-point scale based on frequency, with higher scores indicating greater agitation \citep{Cohen-Mansfield1989}.
The antipsychotic use outcome indicates whether a resident received an antipsychotic medication in the past week.

The trial team also collected information on intervention fidelity. At the individual level, the amount of personalized music played by each resident was recorded using iPod metadata. At the NH level, investigators documented nurse involvement, playlist personalization, and clinical targeting.
Nurse involvement was defined as the proportion of recorded intervention-delivery instances delivered by nursing staff rather than by other NH staff. Playlist personalization was summarized by the proportion of unique songs loaded onto residents' iPods. Clinical targeting was summarized by the clinical proportion, defined as the proportion of enrolled residents in intervention facilities who were identified by NH staff as individuals whose behaviors they intended to target with the personalized music intervention.
A detailed description of the trial design is provided in \citep{McCreedy2021}.

\section{Methods}

\subsection{Notation and potential outcomes}
\label{sec:notation}

We consider a PCRCT with $I$ clusters indexed by
$i=1,\ldots,I$. Individuals within cluster $i$ are indexed by $j=1,\ldots,n_i$, and the total number of participants is $N= \sum_{i=1}^{I}n_i$.
Treatment is assigned at the cluster level, with
$W_i\in\{0,1\}$ indicating if cluster $i$ is assigned to the intervention ($W_i=1$)
or control ($W_i=0$). 

Let $R_{ij}(w)\in\{0,1\}$ denote the potential receipt indicator for individual $j$ in cluster $i$ under cluster assignment $w\in\{0,1\}$, and let $\mathbf{R}_{i}(w) = \{R_{i1}(w),\ldots,R_{i n_i}(w)\}$ denote the corresponding cluster-level receipt vector. Let $\mathbf{Q}_i(w)=\{Q_i^1(w),\ldots,Q_i^L(w)\}$ denote the vector of cluster-level implementation measures under assignment $w$, which are continuous and may be  multivariate.

Potential outcomes are denoted by $Y_{ij}(w,\mathbf{r})$ and represent the outcome for individual $j$
in cluster $i$ under cluster assignment $w$ and cluster receipt vector $\mathbf{r}\in\{0,1\}^{n_i}$. This notation allows for partial interference within clusters, as an individual's outcome may be influenced by the intervention receipt statuses of other individuals in the same cluster \citep{Sobel2006}. Observed data are determined by treatment assignment and satisfy
\begin{align}
Y^{\text{obs}}_{ij}
&= W_i\,Y_{ij}\!\bigl(1,\mathbf{R}_i(1)\bigr)
   + (1-W_i)\,Y_{ij}\!\bigl(0,\mathbf{R}_i(0)\bigr), \\
R^{\text{obs}}_{ij}
&= W_i\,R_{ij}(1) + (1-W_i)\,R_{ij}(0), \\
\mathbf{Q}_i^{\text{obs}}
&= W_i\,\mathbf{Q}_i(1) + (1-W_i)\mathbf{Q}_i(0).
\end{align}
The corresponding missing quantities are
\begin{align}
Y^{\text{mis}}_{ij}
&= (1-W_i)\,Y_{ij}\!\bigl(1,\mathbf{R}_i(1)\bigr)
   + W_i\,Y_{ij}\!\bigl(0,\mathbf{R}_i(0)\bigr), \\
R^{\text{mis}}_{ij}
&= (1-W_i)\,R_{ij}(1) + W_i\,R_{ij}(0), \\
\mathbf{Q}_i^{\text{mis}}
&= (1-W_i)\,\mathbf{Q}_i(1) + W_i\,\mathbf{Q}_i(0).
\end{align}

We further introduce a latent categorical variable
$S_i\in\{1,\ldots,K\}$ indexing unobserved
heterogeneity in intervention implementation and outcomes. The role of $S_i$ and its relationship to observed quantities are specified in
Section~\ref{sec:model}. Throughout, we will refer to $S_i$ as the implementation stratum. For each cluster, we observe $P$ baseline covariates $\mathbf{Z}_i = \{Z_{i}^{1},\ldots, Z_{i}^{P}\}$.
Similarly, for individual $j$ in cluster $i$ we record $M$ pre-treatment covariates $X_{ij} = \{X_{ij}^1,\ldots,X_{ij}^M\}$.

\subsection{Assumptions}\label{sec:Assumptions}

We impose the following assumptions on the data-generating mechanism.
\ \\ 
\noindent \textbf{Assumption 1}: Cluster randomization.
\[
W_i \;\perp\; 
\bigl\{ \mathbf{Y}_{i}(w,\mathbf{r}), \mathbf{R}_{i}(w), \mathbf{Q}_i(w) \bigr\}_{w \in \{0,1\},\, \mathbf{r} \in \{0,1\}^{n_i}}
\;|\; \mathbf{Z}_i,
\]
where $\mathbf{Y}_i(w,\mathbf{r}) =  \{{Y}_{ij}(w,\mathbf{r})\}$ and $\mathbf{Z}_i$ denotes baseline cluster-level covariates (if any) used in randomization.
\ \\ 
\noindent \textbf{Assumption 2}: One-sided noncompliance: $R_{ij}(0) = 0 \  \forall \ i,j$ and $\mathbf{Q}_i(0) = \mathbf{0}$.

Under one-sided noncompliance, clusters assigned to control do not have access to the intervention and individuals cannot receive the intervention unless their cluster is assigned to it. Thus, $R_{ij}(0) = 0 \ \forall \ i,j$, and we define the individual-level compliance indicator
\[
D_{ij} := R_{ij}(1),
\]
so that $D_{ij}=1$ indicates a complier, an individual who would receive the intervention if assigned to treatment, and $D_{ij}=0$ indicates a never-taker. Under this definition, individual receipt satisfies $R_{ij}(w) = w \cdot D_{ij}$, and the vector of potential receipts satisfies $\mathbf{R}_i(w)=w \cdot \mathbf{D}_i$. Similarly, $\mathbf{Q}_i(0) = \mathbf{0}$ because cluster-level implementation does not occur under control assignment. We therefore define the cluster-level implementation vector 
\[
\mathbf{C}_{i} := \mathbf{Q}_{i}(1).
\]
We use collection notation for observed and missing compliance-related quantities: $\mathbf{D}^{\text{obs}} = \{ D_{ij} : W_i = 1 \}$ and $\mathbf{D}^{\text{mis}} = \{ D_{ij} : W_i = 0 \}$ for individual compliance, and $\mathbf{C}^{\text{obs}} = \{ \mathbf{C}_i : W_i = 1 \}$ and $\mathbf{C}^{\text{mis}} = \{ \mathbf{C}_i : W_i = 0 \}$ for cluster-level implementation.

\noindent \textbf{Assumption 3}: No interference across clusters.

Potential outcomes for individual $j$ in cluster $i$ may depend on the assignment $W_i$ and the receipt vector $\mathbf{R}_i(w)$ of that cluster, but do not depend on treatment assignments or receipt statuses of individuals in other clusters. This assumption is common for cluster randomized trials with facilities that are not close to each other geographically, and is implicit in the notation defined in Section~\ref{sec:notation}.

\noindent \textbf{Assumption 4}: No interference given latent implementation strata.

This assumption states that within-cluster interference operates through the latent implementation stratum $S_i$. Under Assumption 2, $\mathbf{R}_i(w) = w\mathbf{D}_i$, so $Y_{ij}(w,\mathbf{r})$ can be written as $Y_{ij}(w, w\mathbf{D}_i)$. We assume that
\[
Y_{ij}(w, w\mathbf{D}_i) = Y_{ij}(w, wD_{ij}, S_i).
\]
Thus, conditional on $S_i$, an individual's potential outcome depends on their own receipt status but not directly on the receipt statuses of other individuals in the cluster. The latent variable $S_i$ summarizes the shared implementation environment, such as staff behavior and intervention delivery practices, that induces dependence among individuals within a cluster.

Under this assumption, potential outcomes and compliance may remain correlated within clusters; however, once $S_i$ is specified, receipt of the intervention by individual $j$ does not directly affect the outcome of another individual $j' \neq j$ in cluster $i$. This formulation can be interpreted as a latent exposure mapping \citep{Manski2013, Aronow2017}, in which dependence of individual outcomes on peers' treatment receipt operates through the unobserved implementation environment $S_i$, rather than through a deterministic function of $\mathbf{D}_i$.

This assumption is plausible in the METRIcAL setting. First, the intervention is delivered at the individual level, with each resident receiving a dedicated music player containing personally curated music, which limits direct interference through intervention receipt. Second, any cross-resident dependence is expected to arise primarily through facility-level implementation processes, such as staffing, workflow, and delivery practices, rather than through one resident's receipt directly changing another resident's outcome. Such shared processes are represented by the latent implementation stratum $S_i$.

Randomized trials with noncompliance often impose an exclusion restriction \citep{Angrist1996}, which rules out effects of assignment on outcomes for noncompliers. We impose a weaker stochastic version that requires this restriction to hold in expectation.
\ \\
\noindent \textbf{Assumption 5}: Stochastic exclusion restriction for noncompliers \citep{Bible_of_Causal}.

For individuals who would not receive the intervention when assigned to treatment,
\[
\mathbb{E}\!\left[ Y_{ij}(1,0, S_i) - Y_{ij}(0,0,S_i)
\;\middle|\;
D_{ij}=0,\, X_{ij}, \mathbf{Z}_i, \mathbf{C}_i, S_i \right] = 0.
\]
Under one-sided noncompliance, $R_{ij}(0)=0$, so the relevant control potential outcome is $Y_{ij}(0,0,S_i)$. This exclusion restriction is in expectation rather than pointwise, and implies that conditional on observed covariates and latent implementation strata, the potential outcomes for individual noncompliers are the same in expectation under both treatment and control.

This assumption is violated if treatment assignment affects outcomes for noncompliers through pathways other than intervention receipt. A possible mechanism is awareness of assignment status among noncompliers, which could induce behavioral or psychological responses. In METRIcAL, this concern is mitigated by the study population. Participants are nursing home residents with moderate to severe dementia and are unlikely to have sustained awareness of trial participation or assignment status. A second possible mechanism is that NH staff aware of the randomized intervention may alter their behavioral management approach for residents who do not receive it. The stochastic exclusion restriction requires only that such effects average to zero conditional on observed covariates and latent implementation strata, rather than holding for every individual. In METRIcAL, systematic differences in staff response to assignment among noncompliers, beyond those captured by $X_{ij}$, $\mathbf{Z}_i$, $\mathbf{C}_i$, and $S_i$, are expected to be negligible.
\subsection{Causal estimands}
\label{sec:estimands}

Causal estimands summarize the effects of an intervention in a population of interest \citep{Gutman2015, Rubin1978}. The population may be restricted to the observed sample (finite-sample estimands) or extended to a target super-population (super-population estimands). Principal stratification estimands characterize effects within subpopulations defined by post-assignment variables such as treatment receipt.

We define the complier average causal effect within latent implementation stratum $k$ as
\begin{equation}
\label{eq:cace_k}
\text{CACE}_k
\;=\;
\mathbb{E}\!\left[
Y_{ij}(1,1,S_i) - Y_{ij}(0,0,S_i)
\;\middle|\;
D_{ij}=1,\; S_i = k
\right],
\qquad k = 1,\ldots,K.
\end{equation}
This estimand compares outcomes under intervention and control assignment among compliers in latent implementation stratum $k$.  The expectation is taken with respect to the target population of interest. Because $\text{CACE}_k$ depends on the mixture specification for $S_i$, it is interpreted as a model-based summary of effect heterogeneity across latent implementation strata.

A marginal complier average causal effect is obtained by averaging over the distribution of $S_i$ among compliers:
\begin{equation}
\label{eq:cace}
\text{CACE}
\;=\;
\mathbb{E}\!\left[
Y_{ij}(1,1,S_i) - Y_{ij}(0,0,S_i)
\;\middle|\;
D_{ij}=1
\right].
\end{equation}

We also define the intent-to-treat (ITT) effect within latent implementation stratum $k$:
\begin{equation}
\label{eq:itt}
\text{ITT}_k
\;=\;
\mathbb{E}\!\left[
Y_{ij}(1, D_{ij}, S_i) - Y_{ij}(0,0,S_i)
\;\middle|\;
S_i = k
\right],
\end{equation}
and the marginal ITT
\[
\text{ITT}
=
\mathbb{E}\!\left[
Y_{ij}(1, D_{ij}, S_i) - Y_{ij}(0,0,S_i)
\right].
\]

\subsection{Identification considerations}
\label{sec:identification}

When clusters do not affect one another's outcomes, randomization identifies the marginal ITT, but it does not by itself identify the marginal super-population CACE. Under one-sided noncompliance, the ITT is a weighted average of the effects of assignment among compliers and noncompliers. The stochastic exclusion restriction assumes that the average effect of assignment among noncompliers is zero. The marginal CACE is then the marginal ITT divided by the proportion of compliers, given this proportion is nonzero. 

This identification result does not require a model for latent implementation strata. However, randomization alone does not identify effects within these strata because stratum membership is unobserved in both arms. We use the joint Bayesian model to estimate marginal and stratum-specific effects and to predict finite-sample effects by imputing unobserved compliance and potential outcomes. The credible intervals quantify uncertainty conditional on the model and its assumptions.

Appendix~\ref{app:model_details} explains which variables each part of the model depends on and where we assume conditional independence. Section~\ref{sec:model} specifies the distributions used for estimation and imputation. Appendix Table~\ref{tab:identification_summary} distinguishes what can be identified from randomization alone, what requires additional assumptions, and where our analysis uses the joint model for estimation.




\section{Bayesian model}
\label{sec:model}

Model-based estimation is a principled way to impose structure on the joint distribution of observed and unobserved variables and to incorporate covariates flexibly \citep{Bible_of_Causal}. The latent-variable model introduced below could also be estimated using likelihood-based methods. We use a Bayesian formulation because posterior sampling provides a convenient framework for jointly accounting for uncertainty in model parameters and unobserved quantities, including latent implementation strata, when making inferences for the causal estimands. Bayesian approaches also enable assessment of model fit using posterior predictive checks \citep{Mattei2013}. Conditional on the resident-level baseline covariates $\mathbf{X}$, the Bayesian model defines a joint distribution for
$(\mathbf{Y}(1,\mathbf{D}), \mathbf{Y}(0,0), \mathbf{D}, \mathbf{C}, \mathbf{Z}, \mathbf{S}),$
with the observed data arising through randomization and the assumptions in Section~\ref{sec:Assumptions}. Under Assumptions~2 and 4, the potential outcomes reduce to $(\mathbf{Y}(1,\mathbf{D}), \mathbf{Y}(0,0))$, where $\mathbf{Y}(1,\mathbf{D}) = \{Y_{ij}(1, D_{ij}, S_i)\}$ and $\mathbf{Y}(0,0) = \{Y_{ij}(0, 0, S_i)\}$.

Units are considered exchangeable when the joint distribution of their potential outcomes is invariant to permutations of their ordering \citep{Bernardo1996}. The Bayesian model imposes partial exchangeability across clusters and, conditional on cluster-level quantities, across individuals within clusters. Based on these assumptions, the joint distribution can be specified through cluster-level covariates and implementation measures, individual compliance under treatment, potential outcomes conditional on compliance status, covariates, and latent implementation strata.

We impose conditional independence restrictions to simplify the parametric specification of the joint distribution. Specifically, conditional on $\mathbf{S}$ and $\mathbf{X}$, we assume that $\mathbf{D}$, $\mathbf{Y}(1,\mathbf{D})$, and $\mathbf{Y}(0,0)$ are jointly independent of $\mathbf{C}$ and $\mathbf{Z}$. We further assume conditional independence of individuals' potential outcomes under different treatment assignments given the shared cluster-level outcome effects and outcome-model parameters (no contamination of imputation across treatments; \citealp{Rubin2008}). Formal statements of the exchangeability and conditional independence assumptions are provided in Appendix~\ref{app:model_details}. Suppressing the model parameters for readability, the conditional joint distribution factors as
\begin{align}
P(\mathbf{Y}(1,\mathbf{D}),\mathbf{Y}(0,0),\mathbf{D},\mathbf{C},\mathbf{Z},\mathbf{S}\mid\mathbf{X})
&=P(\mathbf{Y}(1,\mathbf{D}),\mathbf{Y}(0,0)\mid \mathbf{D},\mathbf{X},\mathbf{S}) \\
&\quad\times P(\mathbf{D}\mid \mathbf{X},\mathbf{S})
P(\mathbf{C},\mathbf{Z}\mid\mathbf{S})P(\mathbf{S}).\nonumber
\end{align}
We condition on resident covariates when fitting the model. For super-population inference, we average over a pooled or stratum-specific empirical distribution of those covariates, as described in Appendix~\ref{app:superpopDerivation}.
\subsection{Latent implementation strata}
\label{sec:latent_classes}

The variable $S_i \in \{1,\ldots,K\}$ represents qualitatively distinct implementation contexts, reflecting differences in organizational capacity, staff engagement, leadership support, and other unmeasured features that influence implementation measures and individual behavior. Latent implementation strata follow a categorical distribution,
\[
S_i \sim \text{Categorical}(\pi_1,\ldots,\pi_K),
\]
with a Dirichlet prior placed on $(\pi_1,\ldots,\pi_K)$ to reflect prior uncertainty in the prevalence of each implementation stratum.

Following Section \ref{sec:Assumptions}, $\mathbf{C}_i := \mathbf{Q}_i(1) = (C_{i1},\ldots,C_{iL})$ denotes the vector of cluster-level implementation measures that would be observed for cluster $i$ under assignment to the treatment arm. Conditional on $S_i$, baseline cluster-level covariates $\mathbf{Z}_i$ and implementation measures $\mathbf{C}_i$ are modeled as
\[
\begin{pmatrix}
\mathbf{C}_i \\
\mathbf{Z}_i
\end{pmatrix}
\;\bigg|\;
S_i = k
\;\sim\;
\text{MVN}\!\left(
\boldsymbol{\mu}^{S}_{k},
\Sigma
\right),
\qquad k=1,\ldots,K,
\]
where $\boldsymbol{\mu}^{S}_{k}
=
\left(
\mu_{k}^{C_1},\ldots,\mu_{k}^{C_L},
\mu_{k}^{Z_1},\ldots,\mu_{k}^{Z_P}
\right)$ and $\Sigma$ is a covariance matrix. This specification allows observed implementation measures to inform inference on $S_i$, while allowing for correlation between baseline cluster characteristics and implementation behavior within latent strata. Although we adopt a multivariate normal specification for $(\mathbf{C}_i, \mathbf{Z}_i)\mid S_i$, alternative mixture kernels could be used without altering the role of $S_i$.

\subsection{Individual Compliance}
\label{sec:compliance_model}

For treated clusters ($W_i=1$), individual compliance ($D_{ij}$) is observed through $R^{\mathrm{obs}}_{ij}$, whereas for control clusters ($W_i=0$), $D_{ij}$ is unobserved. We model compliance indicators as
\[
D_{ij} \mid S_i = k, X_{ij}, \phi_i^{D}
\sim
\text{Bernoulli}\!\left(
\Phi\!\left(\mu_{k}^{D}+X_{ij}^\top \alpha_{k}+\phi_{i}^{D}
\right)\right),
\]
where $\mu_k^{D}$ is an implementation-stratum-specific baseline propensity to comply, $\alpha_k$ is a vector of individual-level covariate effects on compliance within implementation stratum $k$, and $\phi_i^{D}\overset{iid}{\sim} \mathcal{N}(0,\tau_D^2)$ is a cluster-specific effect representing residual cluster heterogeneity in individual compliance beyond the heterogeneity explained by the latent implementation strata and observed covariates.

\subsection{Outcome model}
\label{sec:outcome_model}

We model the conditional distributions of the potential outcomes $Y_{ij}(w,r,S_i)$ given $S_i$, $D_{ij}$ and $\mathbf{X}_{ij}$. Under one-sided noncompliance, $R_{ij}(w)=wD_{ij}$, so the potential outcomes may be written as $Y_{ij}(w, wD_{ij}, S_i)$. In particular, under control assignment only $Y_{ij}(0,0,S_i)$ is well-defined. All model parameters may be indexed by latent implementation stratum $k$, allowing the distributions of compliance behavior and potential outcomes to vary across latent implementation strata. Conditional on latent implementation stratum, compliance status, covariates, outcome-model parameters, and the shared cluster-specific outcome effect, we assume independence between potential outcomes under treatment and control (\citealp{Rubin2008}). For $k=1,\ldots,K$,
\begin{align}
Y_{ij}\!\bigl(1, D_{ij}, S_i \bigr) \mid S_i=k, D_{ij}, X_{ij}, \phi_i^{Y}
&\sim
\mathcal{N}\!\left(
\mu_k^{Y}
+ X_{ij}^\top\beta_{0,k}
+ D_{ij}\bigl(\delta_{1,k}+X_{ij}^\top\beta_{1,k}\bigr)
+\phi_i^{Y},
\;\sigma_k^2
\right),
\label{eq:Y_treat_model}
\\[6pt]
Y_{ij}(0,0, S_i) \mid S_i=k, D_{ij}, X_{ij}, \phi_i^{Y}
&\sim
\mathcal{N}\!\left(
\mu_k^{Y}
+ X_{ij}^\top\beta_{0,k}
+ D_{ij}\delta_{0,k}
+\phi_i^{Y},
\;\sigma_k^2
\right).
\label{eq:Y_ctrl_model}
\end{align}
The parameter $\mu_k^{Y}$ denotes an implementation-stratum-specific baseline level and $\beta_{0,k}$ is a vector of baseline covariate effects. The parameter $\delta_{0,k}$ represents differences in expected control outcomes between compliers and noncompliers within latent implementation stratum $k$, while $\delta_{1,k}$ is the corresponding difference under treatment. The vector $\beta_{1,k}$ represents variation of the treatment effect with individual-level covariates among compliers. To model between-cluster variability, we assume $\phi_i^{Y} \sim \mathcal{N}(0,\tau_Y^2)$ independently across clusters.

Covariates are included in the joint model for adjustment and posterior imputation of latent and unobserved quantities. We do not interpret regression coefficients as causal effects because we describe causal effects using posterior summaries of the ITT and CACE estimands defined in Section~\ref{sec:estimands}. The proposed framework can also be implemented without baseline covariates by using intercept-only models. In the METRIcAL application, we include baseline cluster-level covariates $\mathbf{Z}_i$ because they are observed in both treatment arms and assist in linking control clusters to treatment-defined implementation strata. Because implementation measures $\mathbf{C}_i$ and individual compliance are not observed under control assignment, omitting $\mathbf{Z}_i$ would reduce the information available for assigning control clusters to treatment-defined implementation strata. Specifically, the posterior implementation-stratum membership for control clusters would be informed primarily by the shared outcome model. We discuss the covariate-free specification in Appendix~\ref{app:covariate_free_model}.

\subsection{Estimation procedures}
\label{sec:procedures}

We describe four procedures to estimate the finite-sample versions of the causal estimands in Section~\ref{sec:estimands}. Estimation procedures~1a and~1b target ITT effects, while procedures~2a and~2b target CACEs. All procedures rely on posterior draws from the model described in Section~\ref{sec:model} using Markov chain Monte Carlo (MCMC) algorithms. We derive a data augmentation algorithm \citep{Tanner1987}, treating latent implementation strata, individual compliance indicators, cluster-level implementation measures, and cluster-specific effects as missing data. The sampling algorithm and conditional distributions are described in Appendix~\ref{app:gibbs}.

Causal estimands are computed using posterior predictive draws of missing potential outcomes. For finite-sample estimands, observed outcomes are used directly. Unobserved potential outcomes are imputed from their posterior predictive distributions. For each posterior draw $m$, we construct completed potential outcomes by combining observed outcomes with imputed counterfactual outcomes:
\[
\tilde{Y}_{ij}^{(m)}(1, D_{ij}^{(m)}, S_i^{(m)}) =
\begin{cases}
Y_{ij}^{\text{obs}}, & \text{if } W_i = 1, \\
Y_{ij}^{(m)}(1, D_{ij}^{(m)}, S_i^{(m)}), & \text{if } W_i = 0,
\end{cases}
\]
\[
\tilde{Y}_{ij}^{(m)}(0, 0, S_i^{(m)}) =
\begin{cases}
Y_{ij}^{\text{obs}}, & \text{if } W_i = 0, \\
Y_{ij}^{(m)}(0, 0, S_i^{(m)}), & \text{if } W_i = 1,
\end{cases}
\]
where $Y_{ij}^{(m)}(\cdot)$ denotes draws from the posterior predictive distribution conditional on the parameters and latent variables in posterior draw $m$.
\ \\
\textbf{Procedure 1a}: The finite-sample ITT effect is estimated at posterior draw $m$, as
\begin{equation}
\widehat{\text{ITT}}^{(m)}
=
\frac{1}{N}
\sum_{i=1}^{I}\sum_{j=1}^{n_i}
\left[
\tilde{Y}_{ij}^{(m)}(1, D_{ij}^{(m)}, S_i^{(m)})
-
\tilde{Y}_{ij}^{(m)}(0, 0, S_i^{(m)})
\right].
\end{equation}
\ \\
\textbf{Procedure 1b}: The $\text{ITT}_k$ is estimated at posterior draw $m$ as
\begin{equation}
\widehat{\text{ITT}}^{(m)}_{k}
=
\frac{
\sum_{i=1}^{I}\sum_{j=1}^{n_i}
\mathbb{I}\!\left\{S_i^{(m)}=k\right\}
\left[
\tilde{Y}_{ij}^{(m)}(1, D_{ij}^{(m)}, S_i^{(m)})
-
\tilde{Y}_{ij}^{(m)}(0, 0, S_i^{(m)})
\right]
}{
\sum_{i=1}^{I}\sum_{j=1}^{n_i}\mathbb{I}\!\left\{S_i^{(m)}=k\right\}
}.
\end{equation}
\ \\
\textbf{Procedure 2a}: The marginal CACE is estimated at each posterior draw $m$ as
\begin{equation}
\widehat{\text{CACE}}^{(m)}
=
\frac{
\sum_{i=1}^I \sum_{j=1}^{n_i}
\mathbb{I}\!\left\{ D_{ij}^{(m)} = 1 \right\}
\left[
\tilde{Y}_{ij}^{(m)}(1, D_{ij}^{(m)}, S_i^{(m)})
-
\tilde{Y}_{ij}^{(m)}(0, 0, S_i^{(m)})
\right]
}{
\sum_{i=1}^I \sum_{j=1}^{n_i}
\mathbb{I}\!\left\{ D_{ij}^{(m)} = 1 \right\}
}.
\end{equation}
\ \\
\textbf{Procedure 2b}: The $\text{CACE}_k$ is estimated at each posterior draw $m$ as
\begin{equation}
\widehat{\text{CACE}}^{(m)}_k
=
\frac{
\sum_{i=1}^I \sum_{j=1}^{n_i}
\mathbb{I}\!\left\{ D_{ij}^{(m)} = 1,\; S_i^{(m)} = k \right\}
\left[
\tilde{Y}_{ij}^{(m)}(1, D_{ij}^{(m)}, S_i^{(m)})
-
\tilde{Y}_{ij}^{(m)}(0, 0, S_i^{(m)})
\right]
}{
\sum_{i=1}^I \sum_{j=1}^{n_i}
\mathbb{I}\!\left\{ D_{ij}^{(m)} = 1,\; S_i^{(m)} = k \right\}
}.
\end{equation}
Posterior summaries (e.g., posterior means and credible intervals) are obtained by aggregating these estimates over the $M$ posterior draws.

\subsubsection{Super-population interpretation}
\label{sec:superpop}

The finite-sample estimands in Section~\ref{sec:procedures} refer to the clusters and residents in the study. Their posterior distributions reflect uncertainty about what was not observed for these residents, including whether residents in control clusters would have complied under treatment and what residents' outcomes would have been under the other assignment.

Super-population estimands instead describe average effects in a target population. At each posterior draw, we use the model parameters to compute these effects under a specified target covariate distribution. Marginal effects also depend on how the latent implementation strata are weighted. Appendix~\ref{app:superpopDerivation} shows how to calculate the super-population ITT and CACE from the model parameters, both marginally and within latent implementation strata.

\section{Simulations}\label{sec:CaseStudies}

We use simulated case studies to examine the operating characteristics of the proposed Bayesian approach under correct model specification and under departures from key components of the proposed joint model. Across all case studies, posterior inference is conducted using the Bayesian model described in Section~\ref{sec:model}. The prior distributions for all model parameters are similar across simulation scenarios and are specified in Appendix~\ref{app:simulation_study}. The data-generating mechanisms (DGMs) that define the true relationship between latent implementation strata, individual compliance behavior, and outcomes vary across case studies. Complete specifications of the DGMs are given in Appendix~\ref{app:simulation_study}.

We consider a cluster randomized design with $I=60$ clusters partitioned into two latent implementation strata. Half of the clusters are assigned to the intervention and half to control.In Case Studies~1--5, each cluster contains 20 individuals. Case Study~6, presented Appendix~\ref{app:simulation_study},
repeats Case Study~1 with unequal cluster sizes. Individual-level baseline covariates include resident age and the Activities of Daily Living (ADL) scale \citep{Morris1999ADL}, with covariate values sampled from Stage~1 of the METRIcAL study. These design choices are motivated by the structure and scale of pragmatic cluster randomized trials in health services research \citep{McCreedy2022, McCreedy2021}. For each simulated dataset, we estimate the posterior distribution of each super-population estimand by evaluating the model-parameter expressions in Appendix~\ref{app:superpopDerivation} at each posterior draw.

\subsection{Case Study 1: Correctly Specified DGMs}

The DGM in this case study is correctly specified relative to the Bayesian model in Section~\ref{sec:model}. It serves as a baseline for assessing posterior calibration and operating characteristics when all modeling assumptions are satisfied. Clusters are assigned to one of two latent implementation strata with equal probability. Conditional on latent implementation strata, the cluster-level implementation measure $C_i$ and one generated cluster-level baseline covariate, $Z_i$, are generated jointly from a two-component multivariate normal mixture model. The two implementation strata have different expectations for the implementation measure and the covariate, and they share a common covariance structure. Across simulation replications, the marginal variances and correlations are varied to assess performance across a range of dependence structures. Individual-level compliance probabilities are generated from a probit model including $\mathbf{S}$, $\mathbf{X}$, and cluster-specific effects. Potential outcomes are generated based on $\mathbf{S}$, $\mathbf{D}$, $\mathbf{X}$ and cluster-specific effects. Both baseline outcome levels and compliance-related treatment effects differ across latent implementation strata. Specifically, the DGM for the observed outcome follows Equations (\ref{eq:Y_treat_model}) and (\ref{eq:Y_ctrl_model}) under the treatment and control arms, respectively. 

\subsection{Misspecified DGMs}

Case Studies~2--5 assess the sensitivity of posterior inference to misspecifications of the proposed model and to weaker separation between latent implementation strata. Each scenario isolates a distinct feature of the DGM while keeping the remaining components similar to the correctly specified case. The goal is to evaluate point-estimate accuracy, posterior coverage, and the size of posterior uncertainty when the modeling assumptions are only approximately satisfied.

\subsubsection{Case Study 2: Misspecified Latent Implementation Strata Model}

This case study examines the model's sensitivity to misspecification of the latent implementation strata mixture model. Cluster-level implementation measures and baseline characteristics are generated from a skewed multivariate $t$ distribution with finite degrees of freedom, inducing heavier tails and asymmetry relative to the multivariate normal mixture in the fitted model. Individual-level compliance behavior and potential outcomes are generated according to the same mechanisms as in Case Study~1. Changes in operating characteristics reflect misspecification of the implementation strata model.

\subsubsection{Case Study 3: Incorrectly Specified Individual Compliance Model}

This case study examines sensitivity to misspecification of the individual-level compliance model. Compliance probabilities are generated using the Burr link function ($\text{Burr}_c(x) = 1-(1+e^x)^{-c}$) with $c=0.5$  \citep{Gutman2013}. This misspecification induces systematic deviations in the relationship between covariates and compliance behavior, producing skewed compliance probabilities concentrated near zero. The latent implementation strata model and outcome-generating mechanisms are the same as those in Case Study~1. This scenario isolates the effect of compliance-model misspecification on posterior inference.

\subsubsection{Case Study 4: Misspecified Outcome Model}

This case study evaluates sensitivity to misspecification of the outcome model. Potential outcomes are generated from a model that includes interactions between individual-level covariates that are omitted from the fitted outcome model. These interactions affect both baseline outcome levels and treatment-related contrasts, inducing heterogeneity that is not included in the proposed model. The latent implementation strata model and individual compliance mechanism are generated as in Case Study~1. This scenario assesses how omission of outcome-generating interactions from the fitted model affects posterior inference.

\subsubsection{Case Study 5: Weak Latent-Stratum Separation}

This case study evaluates sensitivity to weaker separation between the latent implementation strata. The cluster-level implementation measure and baseline covariate are generated from the same two-component multivariate normal mixture model as in Case Study~1, but with closer means. The individual-level compliance and outcome-generating mechanisms are the same as in Case Study~1. This scenario assesses whether posterior inference of stratum-specific estimands remains stable when latent implementation strata are more similar.

\subsection{Results for Simulated Case Studies}

Table~\ref{EstimandTrue} reports the true values of the super-population estimands for Case Studies~1--5. The values are the same for Case Studies~1, 2, and 5 because the individual-level compliance and outcome-generating mechanisms are unchanged. These scenarios differ in the distribution used for the cluster-level variables.

\begin{table}[h] 
{\centering
    
\renewcommand{\arraystretch}{0.8}
\setlength{\tabcolsep}{7pt}
\begin{tabular}{cccc}
\hline 
\multirow{2}{*}{Estimand} & \multicolumn{3}{c}{True Values}\tabularnewline
\cline{2-4} 
 & Case Studies 1, 2, and 5 & Case Study 3 & Case Study 4\tabularnewline
\hline 
$\text{ITT}^{sp}$ & 2.53  & 1.52 & 2.62\tabularnewline
\hline 
$\text{ITT}_{1}^{sp}$ & 2.07 & 1.26 & 2.12\tabularnewline
\hline 
$\text{ITT}_{2}^{sp}$ & 3.00  & 1.78 & 3.12\tabularnewline
\hline 
$\text{CACE}^{sp}$ & 4.39  & 4.37 & 4.54\tabularnewline
\hline 
$\text{CACE}_{1}^{sp}$ & 4.11  & 4.17 & 4.22\tabularnewline
\hline 
$\text{CACE}_{2}^{sp}$ & 4.60  & 4.52 & 4.78\tabularnewline
\hline 
\end{tabular}

\begin{minipage}{12.2cm}\linespread{0.8}
\footnotesize

\end{minipage}

\par}

 \linespread{1.0}
     \belowtablecaption{Super-population estimand values for Case Studies~1--5. Case Study~1 uses the correctly specified model; Case Study~2 uses a skewed-$t$ mixture; Case Study~3 uses a Burr compliance link; Case Study~4 adds outcome interactions; and Case Study~5 weakens latent-stratum separation. Case Studies~1, 2, and 5 share the same true estimand values. Subscripts denote implementation-strata-specific estimands, and the superscript $\mathit{sp}$ indicates that estimands are defined with respect to the super-population.}
\label{EstimandTrue}
\end{table}

Across 2{,}500 replications of Case Study~1, posterior credible intervals achieve approximately nominal coverage for all estimands (Table~\ref{tab:case1_results}). Bias and standardized bias are small, and the ratios of the median posterior SD to the empirical SD are close to one, indicating that typical posterior uncertainty is similar to the empirical variability of the posterior means under correct model specification.

\begin{table}[h]
{\centering
\renewcommand{\arraystretch}{0.75}
\setlength{\tabcolsep}{6pt}
\begin{tabular}{lcccccc}
\hline
Estimand & Coverage (\%) & Bias & StdBias & IW & RMSE & Med. SD Ratio \\
\hline
$\text{ITT}^{sp}$      & 95 & 0.19 & 0.08 & 2.38 & 0.61 & 1.04 \\
$\text{ITT}^{sp}_{1}$  & 96 & 0.16 & 0.08 & 3.18 & 0.75 & 1.09 \\
$\text{ITT}^{sp}_{2}$  & 95 & 0.22 & 0.07 & 3.42 & 0.88 & 1.00 \\
$\text{CACE}^{sp}$     & 95 & 0.35 & 0.08 & 3.91 & 1.01 & 1.03 \\
$\text{CACE}^{sp}_{1}$ & 97 & 0.34 & 0.08 & 6.02 & 1.42 & 1.09 \\
$\text{CACE}^{sp}_{2}$ & 95 & 0.35 & 0.08 & 4.98 & 1.29 & 0.99 \\
\hline
\end{tabular}

\begin{minipage}{12.2cm}
\footnotesize
Coverage: empirical coverage of the 95\% posterior credible interval.  
StdBias: bias divided by the absolute value of the true estimand.  
IW: mean 95\% credible interval width.  
RMSE: root mean squared error.
\newline
Med. SD Ratio: median posterior standard deviation across simulation replications divided by the empirical standard deviation of the posterior means across simulation replications.
\end{minipage}

\par}
\belowtablecaption{Operating characteristics of the proposed method under correct model specification (Case Study 1) across 2{,}500 replications.}
\label{tab:case1_results}
\end{table}

Table~\ref{tab:misspec_results} summarizes operating characteristics under the misspecified DGM and weak-separation settings. In Case Study~2, operating characteristics remain similar to those under correct specification, with small standardized bias, near-nominal coverage, and median SD ratios close to one. In Case Study~3, compliance-model misspecification increases standardized bias for both ITT and CACE estimands. Coverage is at or above 95\% for all six estimands, and median SD ratios above one indicate that typical posterior SDs exceed the empirical SDs of the posterior means. In Case Study~4, outcome-model misspecification increased bias, and coverage fell modestly below the nominal 95\% level for both the marginal ITT (92\%) and marginal CACE (91\%). In Case Study~5, weaker latent-stratum separation has limited impact on the point estimate and coverage of the marginal ITT and CACE but there is an increase in the posterior interval widths for stratum-specific estimands, consistent with their dependence on the estimation of latent implementation strata which is less certain.

\begin{table}[!htbp]
{\centering
\renewcommand{\arraystretch}{0.7}
\setlength{\tabcolsep}{5pt}
\begin{tabular}{lcccccc}
\hline
\multicolumn{7}{c}{\textbf{Case Study 2: Skewed-$t$ Mixture}} \\
\hline
Estimand & Coverage (\%) & Bias & StdBias & IW & RMSE & Med. SD Ratio \\
\hline
$\text{ITT}^{sp}$      & 96 & 0.20 & 0.08 & 2.39 & 0.59 & 1.07 \\
$\text{ITT}^{sp}_{1}$  & 97 & 0.15 & 0.07 & 3.18 & 0.74 & 1.10 \\
$\text{ITT}^{sp}_{2}$  & 95 & 0.23 & 0.08 & 3.41 & 0.90 & 0.99 \\
$\text{CACE}^{sp}$     & 96 & 0.35 & 0.08 & 3.92 & 0.98 & 1.08 \\
$\text{CACE}^{sp}_{1}$ & 97 & 0.30 & 0.07 & 6.03 & 1.40 & 1.10 \\
$\text{CACE}^{sp}_{2}$ & 95 & 0.38 & 0.08 & 4.97 & 1.31 & 0.99 \\
\hline
\multicolumn{7}{c}{\textbf{Case Study 3: Burr Compliance Link}} \\
\hline
$\text{ITT}^{sp}$      & 96 & 0.26 & 0.17 & 1.83 & 0.48 & 1.15 \\
$\text{ITT}^{sp}_{1}$  & 98 & 0.20 & 0.16 & 2.33 & 0.51 & 1.26 \\
$\text{ITT}^{sp}_{2}$  & 96 & 0.31 & 0.18 & 2.66 & 0.70 & 1.06 \\
$\text{CACE}^{sp}$     & 96 & 0.72 & 0.16 & 4.98 & 1.29 & 1.16 \\
$\text{CACE}^{sp}_{1}$ & 98 & 0.64 & 0.15 & 7.34 & 1.59 & 1.29 \\
$\text{CACE}^{sp}_{2}$ & 95 & 0.78 & 0.17 & 6.42 & 1.69 & 1.07 \\
\hline
\multicolumn{7}{c}{\textbf{Case Study 4: Outcome Model Misspecification}} \\
\hline
$\text{ITT}^{sp}$      & 92 & 0.46 & 0.18 & 2.13 & 0.63 & 1.27 \\
$\text{ITT}^{sp}_{1}$  & 96 & 0.41 & 0.20 & 2.88 & 0.70 & 1.28 \\
$\text{ITT}^{sp}_{2}$  & 95 & 0.50 & 0.16 & 3.15 & 0.82 & 1.21 \\
$\text{CACE}^{sp}$     & 91 & 0.80 & 0.18 & 3.42 & 1.05 & 1.28 \\
$\text{CACE}^{sp}_{1}$ & 96 & 0.83 & 0.20 & 5.36 & 1.31 & 1.32 \\
$\text{CACE}^{sp}_{2}$ & 94 & 0.78 & 0.16 & 4.53 & 1.22 & 1.20 \\
\hline
\multicolumn{7}{c}{\textbf{Case Study 5: Weak Latent-Stratum Separation}} \\
\hline
$\text{ITT}^{sp}$      & 96 & 0.20 & 0.08 & 2.47 & 0.62 & 1.07 \\
$\text{ITT}^{sp}_{1}$  & 98 & 0.18 & 0.09 & 4.11 & 0.79 & 1.15 \\
$\text{ITT}^{sp}_{2}$  & 97 & 0.19 & 0.06 & 4.40 & 0.98 & 0.99 \\
$\text{CACE}^{sp}$     & 96 & 0.36 & 0.08 & 4.05 & 1.02 & 1.07 \\
$\text{CACE}^{sp}_{1}$ & 98 & 0.31 & 0.08 & 7.86 & 1.47 & 1.15 \\
$\text{CACE}^{sp}_{2}$ & 96 & 0.32 & 0.07 & 6.87 & 1.42 & 1.00 \\
\hline
\end{tabular}

\begin{minipage}{12.2cm}
\footnotesize
StdBias is defined as bias divided by the absolute value of the true estimand.
\newline
Med. SD Ratio is defined as the median posterior standard deviation across simulation replications divided by the empirical standard deviation of the posterior means across simulation replications.
\end{minipage}

\par}
\belowtablecaption{Operating characteristics under misspecified and weak-separation data-generating mechanisms (Case Studies 2--5) across 2{,}500 replications.}
\label{tab:misspec_results}
\end{table}

We also repeated Case Study~1 with unequal cluster sizes of 10 and 30, keeping the total at 60 clusters and 1{,}200 residents. Coverage and bias were similar to those in Case Study~1, with slightly wider credible intervals (Appendix Table~\ref{tab:case6_results}).

Overall, the simulations demonstrate that the proposed Bayesian approach has approximately nominal posterior coverage and small bias under correct specification, and remains stable under misspecification of the cluster-level mixture distribution. However, under compliance- and outcome-model misspecification, median SD ratios exceed one, showing that near-nominal coverage can correspond to posterior uncertainty that exceeds the empirical variability of the posterior means. Bias is largest under compliance- and outcome-model misspecification, while weaker separation between latent strata increases uncertainty for stratum-specific estimands. Code used to produce these simulated case studies can be found in the Supplementary Materials.

\section{Application to METRIcAL data}
\label{DataAnalysis}

We apply the proposed method to METRIcAL to estimate super-population ITT and CACE of the personalized music intervention on change in CMAI from baseline to 4-month follow-up and antipsychotic use at 4-month follow-up. Latent implementation strata are represented using $K=2$ latent classes informed by one baseline facility characteristic, average daily census, and one treatment-arm implementation measure, the clinical proportion.

A NH's average daily census describes its number of housed residents. This covariate is a measure of facility size that can reflect staffing patterns and workflow, as well as the feasibility of consistently delivering personalized music. The clinical proportion is an implementation measure that captures the intervention's intended reach within each facility. It is the proportion of enrolled residents in intervention facilities whose behaviors NH staff intended to address with personalized music as specified on a pre-treatment survey. Other facility characteristics and implementation fidelity measures were collected, but did not reveal substantial separation across facilities. Given the modest number of facilities, we therefore opted for a parsimonious model. Facility-level summaries of enrolled resident characteristics may also be incorporated as cluster-level covariates. We opted to use LTCfocus measures because they describe the broader nursing home population rather than only enrolled study participants.

A resident is considered a complier if they used the intervention for an average of at least 30 minutes per day during the four months following initiation, in accordance with the trial protocol \citep{McCreedy2022}. The CACE estimands are therefore threshold-specific effects among residents who would meet this average-use definition when assigned to the intervention. Baseline resident covariates include MDS-derived variables and baseline CMAI. Six residents had missing baseline covariate values, which were mean imputed because the amount of missing baseline covariate information was small and did not materially affect the analysis. Missing outcome values were assumed to be missing at random and were sampled from their posterior predictive distributions within the Gibbs sampler. For super-population inference, we average over the empirical distribution of resident covariates within each latent implementation stratum (Appendix~\ref{app:superpopDerivation}).

\subsection{CMAI Outcome}
\label{CMAIOutcome}

We analyze change in Cohen--Mansfield Agitation Inventory (CMAI) score from baseline to 4-month follow-up as a continuous outcome. Inference targets super-population ITT and CACE effects, marginally and within latent implementation strata. For residents in facilities belonging to latent implementation stratum $k$, the potential outcomes were modeled as
\begin{align*}
Y_{ij}(1,D_{ij}, S_i) \mid S_i=k, D_{ij}, X_{ij}, \phi_i^{Y}
&\sim
\mathcal{N}\!\left(
\mu_k^{Y}
+ X_{ij}^\top \beta_0
+ D_{ij}\bigl(X_{ij}^\top \beta_1 + \delta_{1,k}\bigr)
+ \phi_i^{Y},
\;\sigma^2
\right) \\
Y_{ij}(0,0, S_i) \mid S_i=k, D_{ij}, X_{ij}, \phi_i^{Y}
&\sim
\mathcal{N}\!\left(
\mu_k^{Y}
+ X_{ij}^\top \beta_0
+ D_{ij}\,\delta_{0,k}
+ \phi_i^{Y},
\;\sigma^2
\right)
\end{align*}
where baseline covariate effects ($\beta_0$), compliance--covariate interactions under treatment ($\beta_1$), and the outcome variance ($\sigma^2$) are shared across latent implementation strata. The
intercept $\mu_k^{Y}$ and compliance-related coefficients $\delta_{0,k}$ and $\delta_{1,k}$ vary across latent implementation strata, allowing baseline outcome levels and treatment-related contrasts to differ across implementation strata. The latent implementation stratum for each NH is inferred from a finite Normal mixture model with $K=2$. Individual compliance is modeled using the multilevel probit specification in Section~\ref{sec:compliance_model}. Prior distributions and additional implementation details are provided in Appendix~\ref{app:priors_application}. For each outcome analysis, we used 4 MCMC chains with 3{,}500 iterations per chain, discarding the first 500 iterations as burn-in and retaining every third subsequent draw to obtain 1{,}000 posterior draws per chain. To address the possibility of label switching \citep{Stephens2000}, at each retained posterior draw, the latent implementation strata were ordered by the first component of the mixture mean vector. The posterior difference between these component means was clearly separated from zero, supporting this ordering rule for consistent labeling. MCMC diagnostics indicated good mixing, with Gelman--Rubin statistics below 1.1 for all reported ITT and CACE estimands. Additional computational diagnostics, including effective sample sizes and trace plots, as well as details on the post-processing relabeling procedure and a sensitivity analysis for the number of latent implementation strata are provided in Appendix~\ref{app:Application}.

\begin{figure}[h]
 \centering
 \includegraphics[width =0.65\linewidth]{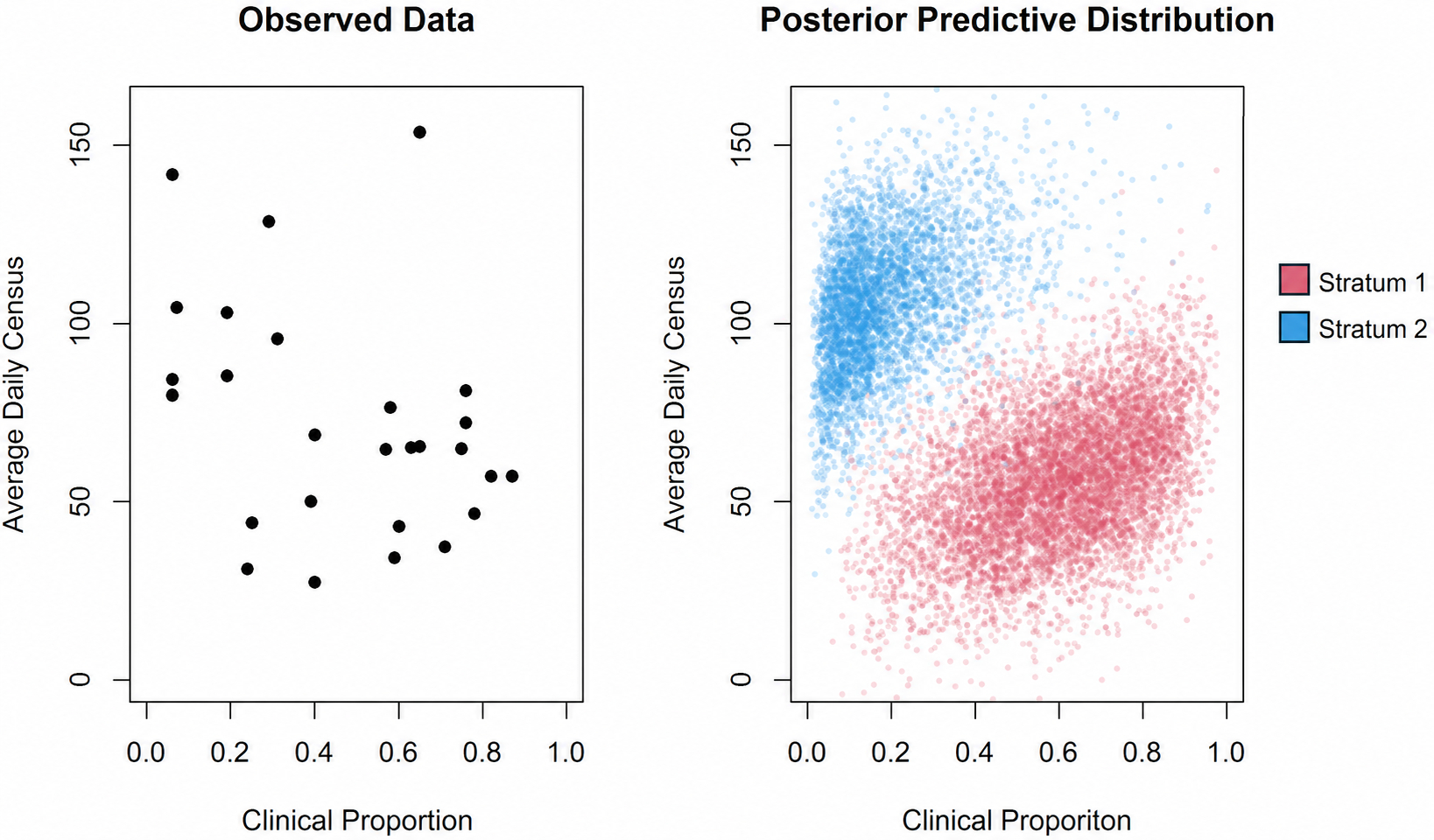} 
\vspace{-10pt}
\linespread{1.0}
    \caption{Observed facility-level data and posterior predictive draws from the fitted mixture model for the clinical proportion (facility-level implementation measure) and average daily census. Posterior predictive draws are colored red for latent implementation stratum 1 and blue for latent implementation stratum 2.}
    \label{PPmix}
\end{figure}

Figure~\ref{PPmix} illustrates the posterior predictive distribution of the clinical proportion and average daily census under the fitted mixture model alongside the observed facility-level data. The posterior predictive draws indicate separation between the two latent implementation strata, with stratum~1 facilities characterized by smaller average daily census and higher clinical proportions relative to stratum~2 facilities. This pattern suggests systematic differences in how facilities operationalized the personalized music intervention, with smaller facilities more likely to target a larger fraction of enrolled residents.

\begin{table}[h]  
{\centering
\renewcommand{\arraystretch}{1}
\setlength{\tabcolsep}{5pt}
\begin{tabular}{ccccc}
\hline 
Estimand & Treatment & Control & $\textbf{Difference}$ & $\textbf{95\% Cr.I}$\tabularnewline
\hline 
\hline 
$\text{ITT}^{sp}$  & -0.45 (0.82)$^\dagger$ & -1.58 (0.76) & 1.13 (0.85) & {[}-0.55, 2.86{]}\tabularnewline
\hline 
 $\text{ITT}^{sp}_{1}$ & -1.87 (0.95) & -2.69 (0.96) & 0.82 (0.93) & {[}-0.95, 2.73{]}\tabularnewline
\hline 
$\text{ITT}^{sp}_{2}$ & 1.78 (1.42) & 0.15 (1.34) & 1.63 (1.58) & {[}-1.51, 4.84{]}\tabularnewline
\hline 
 $\text{ITT}^{sp}_{1}-\text{ITT}^{sp}_{2}$ & - & - & -0.82 (1.83) & {[}-4.38, 2.74{]}\tabularnewline
\hline 
$\text{CACE}^{sp}$ & -0.25 (1.48) & -3.22 (1.96) & 2.98 (2.21) & {[}-1.49, 7.29{]}\tabularnewline
\hline 
$\text{CACE}^{sp}_{1}$  & -2.30 (1.82) & -4.77 (2.37) & 2.46 (2.78) & {[}-2.98, 7.78{]}\tabularnewline
\hline 
$\text{CACE}^{sp}_{2}$  & 2.08 (2.16) & -1.48 (2.99) & 3.56 (3.42) & {[}-3.37, 10.20{]}\tabularnewline
\hline 
$\text{CACE}^{sp}_{1}-\text{CACE}^{sp}_{2}$ & - & - & -1.10 (4.38) & {[}-9.57, 7.45{]}\tabularnewline
\hline 
\end{tabular}
\begin{minipage}{13cm}\linespread{1.0}
\footnotesize
$\dagger$ Posterior Mean (SD)
\end{minipage}
\par}
\linespread{1.0}\belowtablecaption{Posterior distribution of super-population estimands when the outcome is change in CMAI from baseline to 4-month follow-up. $\text{ITT}_{k}$ refers to the intent-to-treat effect among residents in NHs with implementation stratum $k$; $\text{CACE}_{k}$ refers to the complier average causal effect among residents in NHs with implementation stratum $k$. A resident is considered a complier if they listened to the personalized intervention for at least 30 minutes per day on average during the four months after initiation.}
\label{EstimandPost}
\end{table}

Table~\ref{EstimandPost} summarizes posterior inference for super-population ITT and CACE estimands for 4-month change in CMAI. The Treatment and Control columns summarize posterior mean changes from baseline under each arm, whereas the Difference column reports the treatment-minus-control causal contrast. The posterior mean ITT suggests a smaller reduction in CMAI under the intervention relative to standard care (ITT${}^{sp}$: 1.13; 95\% CrI: [-0.55,\,2.86]). The corresponding CACE is larger in magnitude (CACE${}^{sp}$: 2.98; 95\% CrI: [-1.49,\,7.29]), but with greater uncertainty.

The arm-specific posterior means underlying the ITT estimates suggest that CMAI decreased from baseline under both assignments in stratum~1 (higher clinical targeting and smaller census) and increased under both assignments in stratum~2 (lower clinical targeting and larger census). Within both strata, the treatment-minus-control differences are positive, indicating smaller reductions or larger increases in CMAI under treatment than under control. All 95\% credible intervals for the stratum-specific contrasts include zero.

\subsection{Antipsychotics Outcome}

We examine the effect of the intervention on antipsychotic use at the 4-month follow-up. The outcome is binary, indicating whether a resident received an antipsychotic medication in the past week. Potential outcomes for antipsychotic use are modeled using a Bayesian probit regression that conditions on latent implementation strata, individual compliance status, and baseline covariates. For individuals in cluster $i$ with latent implementation stratum $k$, we assume
\begin{align}
Y_{ij}(1,D_{ij}, S_i) \mid S_i=k, D_{ij}, X_{ij}, \phi_i^{Y} &\sim \text{Bernoulli}\!\left(
\text{probit}^{-1}\!\left(
\mu_k^{Y} + X_{ij}^\top \beta_0
+ D_{ij}\bigl(X_{ij}^\top \beta_1 + \delta_{1,k}\bigr)
+ \phi_i^{Y}
\right)
\right), \\
Y_{ij}(0,0,S_i) \mid S_i=k, D_{ij}, X_{ij}, \phi_i^{Y} &\sim \text{Bernoulli}\!\left(
\text{probit}^{-1}\!\left(
\mu_k^{Y} + X_{ij}^\top \beta_0
+ D_{ij}\delta_{0,k}
+ \phi_i^{Y}
\right)
\right).
\end{align}

Table~\ref{TabantipsychResults} summarizes posterior inference for the super-population estimands, reported as marginal risk differences on the percentage-point scale. The posterior mean of the ITT estimand suggests lower antipsychotic use among residents in nursing homes assigned to the personalized music intervention relative to standard care (ITT$^{sp}$: -2.9 percentage points; 95\% CrI: [-6.1,\,0.3]), although the 95\% credible interval includes zero.

The corresponding complier average causal effect indicates a larger estimated reduction among residents who complied with the intervention (CACE$^{sp}$: -7.8 percentage points; 95\% CrI: [-16.9,\,0.7]). While the posterior distribution favors a reduction in antipsychotic use among compliers, uncertainty remains substantial.

Implementation-stratum-specific estimates suggest heterogeneity in the complier effect. The posterior mean of CACE$^{sp}_1$ is more negative than that of CACE$^{sp}_2$, with an estimated difference of -9.0 percentage points (95\% CrI: [-28.9,\,11.3]). Implementation stratum~1 consists of nursing homes with higher clinical targeting of residents for the personalized music intervention, consistent with the possibility that greater targeting intensity corresponds to larger reductions in antipsychotic use among compliers. Credible intervals for all implementation-stratum-specific effects include zero, so these findings should be interpreted with caution.

\begin{table}[h]  
{\centering
\renewcommand{\arraystretch}{0.75}
\setlength{\tabcolsep}{5pt}
\begin{tabular}{ccccc}
\hline 
Estimand & Treatment & Control & $\textbf{Difference}^\dagger$ & $\textbf{95\% Cr.I}$\tabularnewline
\hline 
\hline 
$\text{ITT}^{sp}$  & 26.7 (1.3) & 29.6 (1.3) & -2.9 (1.6) & {[}-6.1, 0.3{]}\tabularnewline
\hline 
 $\text{ITT}^{sp}_{1}$ & 23.5 (1.6) & 27.3 (2.0) & -3.8 (2.2) & {[}-8.2, 0.4{]}\tabularnewline
\hline 
$\text{ITT}^{sp}_{2}$ & 32.0 (2.3) & 33.4 (2.7) & -1.4 (2.8) & {[}-7.2, 3.6{]}\tabularnewline
\hline 
 $\text{ITT}^{sp}_{1}-\text{ITT}^{sp}_{2}$ & - & - & -2.4 (3.8) & {[}-9.5, 5.5{]}\tabularnewline
\hline 
$\text{CACE}^{sp}$  & 34.0 (3.4) & 41.8 (4.8) & -7.8 (4.4) & {[}-16.9, 0.7{]}\tabularnewline
\hline 
$\text{CACE}^{sp}_{1}$  & 32.1 (4.1) & 44.1 (7.6) & -12.1 (7.3) & {[}-27.3, 1.2{]}\tabularnewline
\hline 
$\text{CACE}^{sp}_{2}$  & 36.5 (4.5) & 39.5 (6.3) & -3.1 (6.2) & {[}-16.0, 8.4{]}\tabularnewline
\hline 
$\text{CACE}^{sp}_{1}-\text{CACE}^{sp}_{2}$ & - & - & -9.0 (10.2) & {[}-28.9, 11.3{]}\tabularnewline
\hline 
\end{tabular}
\begin{minipage}{12.2cm}\linespread{1}
\footnotesize
$^\dagger$ Percentage points
\end{minipage}
\par}
\linespread{1.0}\belowtablecaption{Posterior distribution of super-population estimands for antipsychotic use at 4-month follow-up. $\text{ITT}_{k}$ refers to the average ITT effect among residents in NHs with implementation stratum $k$; $\text{CACE}_{k}$ refers to the complier average causal effect among residents in NHs with implementation stratum $k$. A resident is considered a complier if they listened to the personalized intervention for at least 30 minutes per day on average during the four months after initiation.}
\label{TabantipsychResults}
\end{table}

As a benchmark, conventional randomized-arm ITT analyses yielded estimates similar in direction and magnitude to the model-based marginal ITT estimates. For CMAI change, the unadjusted treatment-minus-control difference was 0.91 (SE 1.17), and the baseline-CMAI-adjusted estimate was 1.49 (SE 0.97). For antipsychotic use, the baseline-antipsychotic-adjusted ITT estimate was -2.9 percentage points (SE 1.7 percentage points). These comparisons indicate that the model-based marginal ITT estimates are broadly consistent with randomized-arm ITT analyses, while the proposed model also provides inference for latent-stratum-specific ITT and CACE estimands.

\subsection{Posterior Predictive Checks}

We assess model fit using posterior predictive checks \citep{Gelman1996, Rubin1984} based on predictive discrepancy measures proposed by \citet{Barnard2003} and implemented by \citet{Mattei2013}. These checks evaluate whether the fitted model can reproduce key features of the observed data that are relevant to the ITT and CACE estimands.

For each posterior draw, let $\mathcal{D}_{w,k}^{\text{study}}(\boldsymbol{\theta})=\{(i,j): S_i=k,\; D_{ij}=1,\; W_i=w\}$ denote the draw-specific set of individual-level compliers in the study data belonging to latent implementation stratum $k$ and assigned to treatment arm $w\in\{0,1\}$. Let $N^{\text{study}}_{w,k}(\boldsymbol{\theta})$ denote the number of individuals in $\mathcal{D}_{w,k}^{\text{study}}(\boldsymbol{\theta})$, and let $\bar{Y}^{\text{study}}_{k}(w;\boldsymbol{\theta})$ and $s^{2,\text{study}}_{w,k}(\boldsymbol{\theta})$ denote the sample mean and variance of the corresponding potential outcomes. We consider three discrepancy measures that summarize signal, noise, and signal-to-noise ratio within each latent implementation stratum:
\[
SI_k^{\text{study}}(\boldsymbol{\theta})
=
\left|
\bar{Y}^{\text{study}}_{k}(1;\boldsymbol{\theta})
-
\bar{Y}^{\text{study}}_{k}(0;\boldsymbol{\theta})
\right|,
\]
\[
NO_k^{\text{study}}(\boldsymbol{\theta})
=
\sqrt{
\frac{s^{2,\text{study}}_{1,k}(\boldsymbol{\theta})}{N^{\text{study}}_{1,k}(\boldsymbol{\theta})}
+
\frac{s^{2,\text{study}}_{0,k}(\boldsymbol{\theta})}{N^{\text{study}}_{0,k}(\boldsymbol{\theta})}
},
\qquad
SINO_k^{\text{study}}(\boldsymbol{\theta})
=
\frac{SI_k^{\text{study}}(\boldsymbol{\theta})}{NO_k^{\text{study}}(\boldsymbol{\theta})}.
\]
These measures capture features of the data that are expected to influence treatment effect contrasts within latent implementation strata.

For each retained posterior draw $\boldsymbol{\theta}^{(m)}$, we compute an observed-data discrepancy, $T^{\text{study}}(\boldsymbol{\theta}^{(m)})$, using the observed outcomes together with the draw-specific latent implementation strata and imputed compliance indicators. We then generate a replicated dataset from the posterior predictive distribution under the same parameter draw and compute the corresponding replicated discrepancy, $T^{\text{rep}}(\boldsymbol{\theta}^{(m)})$, using the same discrepancy function. Posterior predictive $p$-values (PPPVs) are defined as the proportion of retained posterior draws for which the replicated discrepancy is larger than the corresponding observed-data discrepancy,
\[
\text{PPPV}
=
\Pr\!\left(
T^{\text{rep}}(\boldsymbol{\theta})
>
T^{\text{study}}(\boldsymbol{\theta})
\;\middle|\;
\mathbf{Y}^{\text{obs}},\mathbf{D}^{\text{obs}},\mathbf{C}^{\text{obs}},\mathbf{Z},\mathbf{X}
\right).
\]
PPPVs near 0 or 1 indicate systematic discrepancies between observed and model-replicated data, suggesting potential model misspecification that could distort causal contrasts.

Across both CMAI and antipsychotic outcomes, the PPPVs for all discrepancy measures and latent implementation strata range from 0.34 to 0.64 (Table~\ref{PPPVs}). These results suggest that the fitted model reproduces the signal, variability, and signal-to-noise structure observed in the data within each latent implementation stratum, with no evidence of systematic lack of fit in components most relevant for estimating ITT and CACE estimands. Additional posterior predictive checks for compliance rates, outcome means and variances, latent class sizes, and cluster-level implementation summaries are provided in Appendix~\ref{app:Application}. Across these checks, posterior predictive $p$-values did not indicate systematic lack of fit in the model components most directly related to the ITT and CACE estimands.

\begin{table}[h]  
{\centering
\renewcommand{\arraystretch}{0.7}
\setlength{\tabcolsep}{5pt}
\begin{tabular}{lccc}
\hline 
\multicolumn{1}{c}{Outcome} & Signal & Noise & Signal/Noise\tabularnewline
\hline 
\multicolumn{4}{l}{CMAI}\tabularnewline
$\ \ \ S=1$ & 0.62 & 0.47 & 0.62\tabularnewline
$\ \ \ S=2$ & 0.43 & 0.39 & 0.44\tabularnewline
\multicolumn{1}{l}{Antipsychotics} &  &  & \tabularnewline
$\ \ \ S=1$ & 0.61 & 0.39 & 0.64\tabularnewline
$\ \ \ S=2$ & 0.36 & 0.61 & 0.34\tabularnewline
\hline 
\end{tabular}

\par}
\linespread{1.0}\belowtablecaption{Posterior predictive $p$-values (PPPVs) for discrepancy measures summarizing signal ($SI_k$), noise ($NO_k$), and signal-to-noise ratio ($SINO_k$) for CMAI and antipsychotic outcomes across latent implementation strata.}
\label{PPPVs}
\end{table}

\section{Discussion}

We propose a Bayesian method for estimating intervention effects in pragmatic cluster randomized controlled trials (PCRCTs) in which cluster- and individual-level one-sided noncompliance may occur. The method uses a mixture model to infer latent implementation strata from baseline characteristics and continuous implementation measures observed in the intervention arm. Individual-level binary compliance is modeled using a probit regression among individuals in intervention facilities. The joint model is used to impute unobserved cluster-level implementation measures for control facilities and individual-level compliance status for individuals in those facilities. Using these imputed quantities, we estimate ITT and CACE estimands marginally and within latent implementation strata.

In cluster randomized trials, one individual's receipt of the intervention may influence another individual's outcome within the same cluster. Exposure mapping approaches address this possibility by representing interference as a function of observed treatment or receipt patterns within a cluster \citep{Manski2013, Aronow2017}. Instead, our method assumes that, conditional on the latent implementation stratum, within-cluster outcome dependence is driven by shared, partially unobserved implementation processes, such as staff engagement and delivery practices, rather than by peers' receipt directly. These processes are not directly observed and may not be adequately represented as deterministic functions of the observed receipt vector. The latent implementation variable summarizes the shared environment governing compliance behavior and outcome dependence. This can be viewed as a latent analogue of an exposure mapping, in which interference operates through unobserved implementation mechanisms rather than a pre-specified function of peers' treatment receipt.

Implementation measures should be chosen based on the trial protocol and the scientific question defining meaningful implementation heterogeneity. Measures included in $\mathbf{C}_i$ should represent cluster-level features of intervention delivery that are measured consistently across treated clusters and are plausibly related to uptake, intervention fidelity, or other mechanisms of effect heterogeneity. Because latent implementation strata are defined relative to these measures, stratum-specific ITT and CACE estimates should be interpreted conditional on the chosen implementation summaries.

Through simulated case studies, we examined the operating characteristics under correctly specified and misspecified data-generating mechanisms. Under correct specification, estimates had small bias, posterior coverage was approximately nominal, and median SD ratios were close to one. Bias increased under both compliance- and outcome-model misspecification. Coverage remained at or above 95\% under compliance-model misspecification, but under outcome-model misspecification it fell to 92\% for the marginal ITT and 91\% for the marginal CACE. Future simulation work could examine additional departures from the baseline design, including direct assignment effects among noncompliers. Sensitivity analyses for violations of the stochastic exclusion restriction could be conducted by augmenting the outcome model to allow nonzero direct assignment effects among noncompliers, thereby evaluating how CACE estimates change under departures from the current identifying restriction.

We applied the proposed method to METRIcAL, a PCRCT evaluating a personalized music intervention among nursing home residents with dementia across 54 facilities in the United States. The model inferred two latent implementation strata: one characterized by lower average daily census and a higher proportion of residents targeted for the intervention, and the other by higher average daily census and a lower proportion targeted. At the population level, point estimates suggest that residents in intervention facilities experienced smaller reductions in CMAI scores relative to standard care, but also exhibited lower rates of antipsychotic use at four-month follow-up.

Point estimates varied across implementation strata. For CMAI, the arm-specific posterior means underlying the ITT estimates suggest decreases from baseline in the higher-targeting implementation stratum under both treatment and control, and increases from baseline in the lower-targeting stratum under both arms. However, the treatment-minus-control contrasts for CMAI are positive, indicating smaller estimated reductions or larger estimated increases under the intervention than under standard care, with substantial posterior uncertainty. In contrast, antipsychotic use is lower in the intervention arm across both implementation strata, with larger estimated reductions in higher-targeting facilities and among compliers. This pattern suggests that the intervention may be more strongly associated with reduced antipsychotic use than with improved agitation scores, although none of the 95\% credible intervals for the effect estimates excluded zero.

One possible explanation is that behavioral symptoms under standard care may have been managed partly through antipsychotic medications, whereas residents in intervention facilities may have had lower antipsychotic use. This pattern could contribute to smaller reductions or increases in CMAI under the intervention despite potential benefits from personalized music. These findings suggest that the intervention's effectiveness may depend on both facility-level implementation and individual compliance, particularly with respect to reducing reliance on antipsychotic medications. Given the inconclusive nature of the METRIcAL results, future research on music-based interventions for residents with ADRD should consider strategies to improve implementation and compliance, and the potential role of such interventions as alternatives to antipsychotic medications, which have been associated with adverse outcomes in nursing home populations \citep{Huybrechts2012, Chiu2015, Liperoti2017}.

In conclusion, the proposed Bayesian framework provides an approach for analyzing pragmatic CRCTs in which intervention implementation is assessed at the cluster level and individual-level compliance is assessed with a binary indicator. By jointly modeling implementation heterogeneity and individual compliance, the method enables estimation of intervention effects among facilities with similar implementation patterns and among individuals who receive the intervention as intended. Applied to METRIcAL, the approach yields insights not available from standard intent-to-treat analyses alone. Important extensions include models allowing partial or continuous individual-level compliance and more complex interference structures.
\clearpage



\section*{Disclosure Statement}
The authors report no conflict of interest.

\section*{Data Sharing Statement}

Participants could not consent to public data sharing, and the supporting data are not publicly available because of the sensitive nature of the research. Similar synthetic datasets are provided in the supplementary materials.

\section*{Supplementary Material}
\label{sec6}

\textbf{Simulation Code and Analysis Examples.zip} - A folder containing R code and necessary materials to conduct:
\begin{itemize}
    \item The simulated case studies described in this article
    \item Analyses on synthetic data generated using the data analyzed in this article.
    \item Code for a simulation study demonstrating the validity of the derived formulas for the super-population estimands as functions of the model parameters.
\end{itemize}

The supplementary code and data are also available at
{\def\UrlBigBreaks{\do\:\do\-}%
\url{https://github.com/AnthonySisti/Bayesian-Compliance-Modeling-in-Cluster-RCTs-with-One-Sided-Noncompliance}}.

\section*{Acknowledgments}
This work was supported by National Institute on Aging (NIA) grant R21AG057451. 
\newpage
\renewcommand{\bibsection}{\section*{\refname}}

\clearpage
\markboth{Appendix}{Appendix}

\section*{Appendix}

\setcounter{table}{0}
\setcounter{figure}{0}
\setcounter{section}{0}
\setcounter{subsection}{0}
\setcounter{page}{1}
\renewcommand{\thetable}{A\arabic{table}}
\renewcommand{\thefigure}{A\arabic{figure}}
\renewcommand{\thesection}{A\arabic{section}}

\section{Identification and Model-dependence of Causal Estimands}
\label{app:identification_summary}

\begin{table}[!ht]
\centering
\caption{Identification and model-dependence of the causal estimands.}
\label{tab:identification_summary}
\renewcommand{\arraystretch}{1.15}
\begin{tabular}{p{0.27\linewidth}>{\centering\arraybackslash}p{0.10\linewidth}p{0.55\linewidth}}
\hline
Estimand & Design ID? & Additional requirements \\
\hline
Marginal ITT &
Yes &
Cluster randomization and no between-cluster interference. \\

Marginal CACE &
No &
For the super-population mean contrast: the marginal ITT requirements, one-sided noncompliance, the stochastic exclusion restriction for noncompliers, and an identifiable, positive complier proportion under the same resident weighting. The joint model is used for estimation here, but is not required for this identification result. \\

$\text{ITT}_k$ &
No &
Latent implementation-strata model and outcome model, because $S_i$ is latent and is not directly observed for all clusters. \\

$\text{CACE}_k$ &
No &
Latent implementation-strata model, compliance model, outcome model, one-sided noncompliance, and stochastic exclusion restriction. \\

Super-population versions &
Depends &
The relevant requirements above, together with assumptions linking the study sample to the target population. The model-based approach used here additionally specifies the population distributions of covariates, latent strata, compliance, and outcomes. \\
\hline
\end{tabular}
\end{table}

\section{Model Details and Exchangeability Assumptions}
\label{app:model_details}

This appendix formalizes the exchangeability and conditional independence assumptions underlying the Bayesian model described in Section~\ref{sec:model}. These assumptions justify the factorization of the joint distribution of latent and observed variables and clarify the role of latent implementation strata.

Let $\theta_S$ denote the population-level parameters of the latent implementation strata model, including the mixture probabilities, mixture means, and covariance parameters for $(\mathbf{C}_i,\mathbf{Z}_i)\mid S_i$. Let $\theta_D$ denote the population-level parameters of the individual compliance model, including the stratum-specific compliance intercepts, compliance regression coefficients, and the variance component for the compliance effects. Let $\theta_Y$ denote the population-level parameters of the outcome model, including the stratum-specific outcome intercepts, outcome regression coefficients, compliance-related contrasts, outcome variance parameters, and the variance component for the outcome effects. The cluster-specific effects, $\phi_i^D$ and $\phi_i^Y$, are sampled as latent quantities in the augmented posterior.

We assume partial exchangeability at both the cluster and individual levels, consistent with the cluster-randomized design and the hierarchical structure of the model. We condition on resident covariates and treatment assignments when fitting the model.

\begin{enumerate}
\renewcommand{\labelenumi}{(\alph{enumi})}
\item \textbf{Exchangeability of latent implementation strata.}
The latent implementation strata indicators $S_i \in \{1,\ldots,K\}$ are exchangeable across clusters.
Given the mixture proportions $\boldsymbol\pi$, the $S_i$ are independent categorical draws with $\Pr(S_i=k\mid\boldsymbol\pi)=\pi_k$.

\item \textbf{Exchangeability of cluster-level characteristics and implementation measures.}
The cluster-level implementation measures under treatment, $\mathbf{C}_i = \mathbf{Q}_i(1)$, and baseline cluster-level covariates $\mathbf{Z}_i$ are exchangeable across clusters conditional on latent implementation strata.
Given the strata and model parameters, the vectors $T_i=(\mathbf C_i^\top,\mathbf Z_i^\top)^\top$ are drawn independently across clusters from their stratum-specific distributions.

\item \textbf{Partial exchangeability of individual compliance indicators.}
Under one-sided noncompliance, individual-level compliance indicators are defined as
$D_{ij} := R_{ij}(1)$, with $D_{ij}=1$ denoting a complier and $D_{ij}=0$ a never-taker.
We use the same conditional compliance model for residents in the same stratum. Within a cluster, their compliance indicators are independent given the resident covariates, the stratum, the model parameters, and the shared cluster-level compliance effect $\phi_i^D$.

\item \textbf{Partial exchangeability of potential outcomes.}
Under one-sided noncompliance, the relevant potential outcomes are $Y_{ij}(1,D_{ij},S_i)$ and $Y_{ij}(0,0,S_i)$.
Within a cluster, the outcome contributions for different residents are independent given their compliance types and covariates, the stratum, the model parameters, and the shared cluster-level outcome effect $\phi_i^Y$.
\end{enumerate}

{
The cluster-level compliance and outcome effects are independent of
one another and across clusters, conditional on the model parameters.
Let $Y_{ij}^{\mathrm{fact}}$ denote the outcome under the assigned arm,
whether observed or imputed. The complete-data distribution for cluster
$i$ is
\begin{align*}
&P(S_i=k,T_i,\phi_i^D,\phi_i^Y,\mathbf D_i,\mathbf Y_i^{\mathrm{fact}}
       \mid\mathbf X_i,W_i,\boldsymbol\theta)\\
&\quad=\pi_k\,P(T_i\mid S_i=k,\theta_S)\,
            P(\phi_i^D\mid\theta_D)\,P(\phi_i^Y\mid\theta_Y)\\
&\qquad\quad\times\prod_{j=1}^{n_i}
P(D_{ij}\mid X_{ij},S_i=k,\phi_i^D,\theta_D)\\
&\qquad\quad\times\prod_{j=1}^{n_i}
P(Y_{ij}^{\mathrm{fact}}\mid W_i,D_{ij},X_{ij},S_i=k,\phi_i^Y,\theta_Y).
\end{align*}
Here $P$ denotes a density or probability mass function, as appropriate.
If we integrate out the cluster-level effects, we integrate the product
for the entire cluster. Integrating separately for each resident would
remove the dependence induced by the shared effects.
}

In addition, we impose the following conditional independence assumptions.
These are modeling restrictions rather than causal assumptions and may be relaxed in alternative specifications.

\medskip
\noindent
\textbf{Conditional independence from observed cluster-level characteristics.}
Given the latent implementation stratum, resident covariates, and the corresponding cluster-level effects, the compliance and outcome models do not depend directly on the cluster-level characteristics or implementation measures. The outcome model also conditions on individual compliance type. These restrictions give the resident-level factors in the display above.

\medskip
\noindent
\textbf{No contamination of imputation across treatment assignments} \citep{Rubin2008}.
Conditional on latent implementation strata, individual compliance status,
covariates, outcome-model parameters, and the shared cluster-level outcome
effect, the potential outcomes under treatment and control are independent:
\[
Y_{ij}(1,D_{ij}, S_i) \;\perp\; Y_{ij}(0,0, S_i)
\;\big|\;
S_i, D_{ij}, X_{ij}, \phi_i^Y, \theta_Y.
\]
Because $\phi_i^Y$ enters both potential-outcome models,
treatment and control potential outcomes may remain correlated after
integrating over this shared cluster-level effect.

\medskip
\noindent
\textbf{Conditional independence of individual-level covariates.}
For super-population standardization, we specify the target
distribution of resident covariates without further conditioning on
cluster-level characteristics or implementation measures within a stratum.
This distribution can be pooled or stratum specific, as described in
Appendix~\ref{app:superpopDerivation}. It is used to evaluate the estimands,
not as a likelihood factor when fitting the model.

\subsection{Covariate-free specification}
\label{app:covariate_free_model}

The proposed model can be written without baseline covariates by replacing the covariate-dependent components with intercept-only specifications. In this case, latent implementation strata are modeled as
\[
S_i \sim \mathrm{Categorical}(\pi_1,\ldots,\pi_K),
\qquad
\mathbf{C}_i \mid S_i=k \sim \mathrm{MVN}(\boldsymbol{\mu}^C_k,\Sigma_C),
\]
where $\mathbf{C}_i=\mathbf{Q}_i(1)$ denotes implementation measures defined under treatment assignment. The compliance and outcome models reduce to
\[
D_{ij}\mid S_i=k, \phi_i^{D}
\sim
\mathrm{Bernoulli}\!\left[
\Phi\left(\mu_k^D+\phi_i^D\right)
\right],
\]
and
\[
Y_{ij}(w,wD_{ij},S_i)
\mid D_{ij},S_i=k, \phi_i^{Y}
\sim
f_Y\left\{ \eta^Y_{ij}(w,D_{ij},k),\theta_Y \right\},
\]
with the linear predictor containing only stratum-specific intercepts, compliance terms, treatment terms, and cluster-specific effects as appropriate for the continuous or binary outcome model.

This intercept-only specification is a valid reduced version of the joint model, but it changes how latent implementation strata are informed. Because $\mathbf{C}_i$ is only defined and observed for clusters assigned to treatment, control-cluster membership in $S_i$ cannot be informed directly by observed implementation measures. Without baseline cluster-level covariates $\mathbf{Z}_i$ observed in both arms, posterior membership for control clusters is therefore determined primarily through the shared compliance and outcome models and the mixture proportions. Thus, omitting $\mathbf{Z}_i$ does not remove the latent-strata model, but it makes stratum assignment for control clusters more dependent on parametric modeling structure.

\section{Gibbs Sampler Derivation}\label{app:gibbs}

This appendix describes a Gibbs sampler for posterior computation under the general model in Section~\ref{sec:model}, allowing (i) stratum-specific outcome regression coefficients and variances and (ii) stratum-specific individual-level compliance regression coefficients. Throughout, $i=1,\dots,I$ indexes facilities (clusters), $j=1,\dots,n_i$ indexes residents, and $W_i\in\{0,1\}$ is the cluster-level assignment indicator.

The sampler below is written for the general version of the model, in which compliance regression coefficients, outcome regression coefficients, and outcome variances may vary by latent implementation stratum. The simulation study uses this structure closely. In the METRIcAL analysis, we used reduced versions of several blocks to improve stability with the observed number of facilities: compliance slopes were shared across strata with stratum-specific intercepts, and outcome covariate effects were shared across strata with stratum-specific intercepts and compliance-related contrasts. For the binary outcome analysis, the outcome model was fit using the analogous probit data-augmentation step. These are special cases of the Gibbs-sampling framework, with the corresponding design matrices and coefficient blocks modified accordingly.

\subsection*{Augmented posterior}

{
Let $\mathbf O=(\mathbf Y^{obs},\mathbf D^{obs},\mathbf C^{obs},
\mathbf Z,\mathbf X,\mathbf W)$ denote the observed data. Let
$\mathbf C^{mis}$ denote the implementation measures unavailable in
control facilities, and let $\mathbf D^{mis}$ include compliance types
in control facilities and any unrecorded compliance types in treated
facilities. Write $\mathbf Y^{mis}_{\mathrm{fact}}$ for missing outcomes
under the assigned arm, and $\mathbf Y^{\mathrm{fact}}$ for the full
vector of assigned-arm outcomes, observed or imputed. The two vectors
of facility random effects are denoted by $\boldsymbol\phi^D$ and
$\boldsymbol\phi^Y$.

Let $\mathbf L=(\mathbf S,\mathbf D^{mis},\mathbf C^{mis},
\boldsymbol\phi^D,\boldsymbol\phi^Y,\mathbf Y^{mis}_{\mathrm{fact}})$
denote the unobserved strata, missing data, and cluster-level effects.
We obtain the posterior for the model parameters by summing or integrating
over these quantities:
\[
P(\boldsymbol\theta\mid\mathbf O)
=
\sum_{\mathbf S}\sum_{\mathbf D^{mis}}
\int P(\boldsymbol\theta,\mathbf L\mid\mathbf O)\,
d\mathbf C^{mis}\,d\boldsymbol\phi^D\,d\boldsymbol\phi^Y\,
d\mathbf Y^{mis}_{\mathrm{fact}}.
\]
For binary outcomes, the integral over missing outcomes is a sum.
The Gaussian variables used in the probit updates have already been
integrated out.
}

\subsection*{Model components (general form)}

\paragraph{(i) Facility-level mixture for $(C_i,Z_i)$.}
Let $T_i=\begin{bmatrix} C_i \\ Z_i \end{bmatrix}\in\mathbb{R}^q$ where $q=q_C+q_Z$.
\[
S_i\mid \boldsymbol{\pi}\sim \text{Categorical}(\pi_1,\dots,\pi_K),\qquad
T_i\mid S_i=k,\boldsymbol{\mu}^S_k,\Sigma \sim \text{MVN}(\boldsymbol{\mu}^S_k,\Sigma).
\]

\paragraph{(ii) Compliance model (probit, stratum-specific coefficients).}
Let $x_{ij}\in\mathbb{R}^p$ denote resident covariates and define $\eta^D_{ij,k}=\mu^D_k + x_{ij}^\top \alpha_k + \phi^D_i$.
\[
D_{ij}\mid S_i=k,\mu^D_k,\alpha_k,\phi_i^D \sim \text{Bernoulli}\bigl(\Phi(\eta^D_{ij,k})\bigr),
\qquad
\phi^D_i\sim N(0,\tau_D^2).
\]

\paragraph{(iii) Continuous outcome model (stratum-specific coefficients and variances).}
Let $\varepsilon_{ij}\sim N(0,\sigma_k^2)$ when $S_i=k$ and define
\[
Y^{\mathrm{fact}}_{ij}\mid S_i=k, D_{ij},W_i, x_{ij}, \phi_i^Y
\sim
N\!\left(m_{ij,k}(D_{ij}),\ \sigma_k^2\right),
\qquad
\phi^Y_i\sim N(0,\tau_Y^2),
\]
where
\[
m_{ij,k}(D)
=
\mu^Y_k
+ x_{ij}^\top \beta_{0,k}
+ (1-W_i)\,D\,\delta_{0,k}
+ W_i\,D\left(x_{ij}^\top \beta_{1,k}+\delta_{1,k}\right)
+ \phi_i^Y.
\]
(For the antipsychotic outcome, replace the normal likelihood with a probit likelihood and use an analogous Albert--Chib augmentation step as in the compliance model below.)

\subsection*{Prior distributions (general form)}
We assume the priors used in the main paper are special cases of:
\begin{align*}
\boldsymbol{\pi}&\sim \text{Dirichlet}(\lambda_1,\dots,\lambda_K),\\
\boldsymbol{\mu}^S_k&\sim \text{MVN}(m_{\mu^S},V_{\mu^S})\quad (k=1,\dots,K),\qquad
\Sigma\sim \text{Inv-Wishart}(\Psi,\nu_\Sigma),\\
\mu^D_k&\sim N(0,v_{\mu^D}^2),\qquad \alpha_k\sim \text{MVN}(0,V_{\alpha})\quad (k=1,\dots,K),\\
\tau_D&\sim \text{Uniform}(0,U_D),\qquad \tau_Y\sim \text{Uniform}(0,U_Y),\\
\mu^Y_k&\sim N(0,v_{\mu^Y}^2),\qquad
\delta_{0,k}\sim N(0,v_{\delta_0}^2),\qquad \delta_{1,k}\sim N(0,v_{\delta_1}^2)\quad (k=1,\dots,K),\\
\beta_{0,k}&\sim \text{MVN}(0,V_{\beta_0}),\qquad \beta_{1,k}\sim \text{MVN}(0,V_{\beta_1})\quad (k=1,\dots,K),\\
\sigma_k^2&\sim \text{Inv-Gamma}(a_\sigma,b_\sigma)\quad (k=1,\dots,K).
\end{align*}
The uniform priors on $\tau_D$ and $\tau_Y$ imply $P(\tau^2)\propto (\tau^2)^{-1/2}\,\mathbb{I}(0<\tau^2<U^2)$.

\subsection*{Gibbs sampler}

Initialize $(\boldsymbol\theta^{(0)},\mathbf L^{(0)})$.
For $m=1,\dots,M$, use the steps below in the order shown. This order
updates the random effects and their variances before the corresponding
regression coefficients, as in the simulation code.

\subsubsection*{1. Update mixture parameters $(\{\mu_k^S\},\Sigma,\pi)$ given $(\mathbf{S},\mathbf{C},\mathbf{Z})$}

Let $T_i^{(m-1)}=\begin{bmatrix} C_i^{(m-1)} \\ Z_i \end{bmatrix}$, where $C_i^{(m-1)}=C_i$ if $W_i=1$ and $C_i^{(m-1)}=C_i^{mis(m-1)}$ if $W_i=0$.

\paragraph{1(a) $\boldsymbol{\mu}^S_k$ (conjugate MVN).}
Let $n_k=\sum_{i=1}^I \mathbb{I}(S_i^{(m-1)}=k)$,
$t_k=\sum_{i:S_i^{(m-1)}=k}T_i^{(m-1)}$ and
$Q_\Sigma=(\Sigma^{(m-1)})^{-1}$.
Then
\[
\boldsymbol{\mu}^{S(m)}_k \mid \cdot \sim \text{MVN}(M_{\mu,k},\Omega_{\mu,k}),
\]
with
\[
\Omega_{\mu,k}=\left(V_{\mu^S}^{-1}+n_k Q_\Sigma\right)^{-1},
\qquad M_{\mu,k}=\Omega_{\mu,k}\left(V_{\mu^S}^{-1}m_{\mu^S}+Q_\Sigma t_k\right).
\]
For an empty component, $n_k=0$ and $t_k=\mathbf 0$,
so the full conditional is the prior.

\paragraph{1(b) $\Sigma$ (conjugate Inv-Wishart).}
Let
\[
S_\Sigma=\sum_{k=1}^K\sum_{i:S_i^{(m-1)}=k}\left(T_i^{(m-1)}-\boldsymbol{\mu}_k^{S(m)}\right)\left(T_i^{(m-1)}-\boldsymbol{\mu}_k^{S(m)}\right)^\top.
\]
Then
\[
\Sigma^{(m)}\mid\cdot \sim \text{Inv-Wishart}\bigl(\Psi+S_\Sigma,\ \nu_\Sigma+I\bigr).
\]

\paragraph{1(c) $\boldsymbol{\pi}$ (conjugate Dirichlet).}
\[
\boldsymbol{\pi}^{(m)}\mid\cdot \sim \text{Dirichlet}\!\left(\lambda_1+n_1,\dots,\lambda_K+n_K\right).
\]

\subsubsection*{2. Update compliance model parameters $(\{\mu_k^D,\alpha_k\},\phi^D,\tau_D^2)$}

We use Albert--Chib augmentation. Introduce latent variables $U_{ij}$ such that
\[
U_{ij}\mid S_i=k,\mu_k^D,\alpha_k,\phi_i^D \sim N(\eta^D_{ij,k},1),
\qquad
D_{ij}=\mathbb{I}(U_{ij}>0).
\]

\paragraph{2(a) Sample $U_{ij}$.}
For each $(i,j)$, let $k=S_i^{(m-1)}$ and $\eta=\mu_k^{D(m-1)}+x_{ij}^\top \alpha_k^{(m-1)}+\phi_i^{D(m-1)}$.
Then
\[
U_{ij}^{(m)}\mid\cdot \sim
\begin{cases}
N(\eta,1)\ \text{truncated to }(0,\infty), & D_{ij}^{(m-1)}=1,\\
N(\eta,1)\ \text{truncated to }(-\infty,0], & D_{ij}^{(m-1)}=0.
\end{cases}
\]

\paragraph{2(b) Sample $\phi_i^D$ (conjugate normal).}
Let $k=S_i^{(m-1)}$ and define residuals
\[
r_{ij}^D = U_{ij}^{(m)}-\mu_k^{D(m-1)}-x_{ij}^\top \alpha_k^{(m-1)}.
\]
Then
\[
\phi_i^{D(m)}\mid\cdot \sim N\!\left(M_{\phi_i^D},V_{\phi_i^D}\right),
\quad
V_{\phi_i^D}=\left(n_i+1/\tau_D^{2(m-1)}\right)^{-1},
\quad
M_{\phi_i^D}=V_{\phi_i^D}\sum_{j=1}^{n_i} r_{ij}^D.
\]

\paragraph{2(c) Sample $\tau_D^2$ (truncated Inv-Gamma).}
Using $\phi_i^D\sim N(0,\tau_D^2)$ and $\tau_D\sim \text{Unif}(0,U_D)$,
\[
\tau_D^{2(m)}\mid\cdot \sim \text{Inv-Gamma}\!\left(a_D,\ b_D\right)\ \text{truncated to }(0,U_D^2),
\]
where
\[
a_D=\frac{I-1}{2},\qquad b_D=\frac{1}{2}\sum_{i=1}^I \left(\phi_i^{D(m)}\right)^2.
\]

\paragraph{2(d) Sample $(\mu_k^D,\alpha_k)$ (conjugate MVN given $U$).}
For each $k$, stack all $(i,j)$ such that $S_i^{(m-1)}=k$.
Let $N_k=\sum_{i:S_i^{(m-1)}=k} n_i$.
Define the $N_k\times(1+p)$ design matrix $X_k^*=[\mathbf{1}\ \ X_k]$ and response vector
\[
\tilde{U}_k = U_k^{(m)} - \tilde{\phi}_k^{D(m)},
\]
where $\tilde{\phi}_k^{D(m)}$ stacks $\phi_i^{D(m)}\mathbf{1}_{n_i}$ for all $i$ with $S_i^{(m-1)}=k$.
Let $\theta_k^D=\begin{bmatrix}\mu_k^D\\ \alpha_k\end{bmatrix}$, prior $\theta_k^D\sim \text{MVN}(0,V_D)$ with $V_D=\text{blockdiag}(v_{\mu^D}^2,\ V_\alpha)$.
Then
\[
\theta_k^{D(m)}\mid\cdot \sim \text{MVN}(M_{D,k},\Omega_{D,k}),
\]
with
\[
\Omega_{D,k}=\left(X_k^{*^\top}X_k^*+V_D^{-1}\right)^{-1},\qquad
M_{D,k}=\Omega_{D,k}\,X_k^{*^\top}\tilde{U}_k.
\]

\subsubsection*{3. Update outcome model parameters $(\{\mu_k^Y,\beta_{0,k},\delta_{0,k},\beta_{1,k},\delta_{1,k}\},\sigma_k^2,\phi^Y,\tau_Y^2)$}

For each $k$, define the coefficient vector
\[
B_k=\begin{bmatrix}\mu_k^Y\\ \beta_{0,k}\\ \delta_{0,k}\\ \beta_{1,k}\\ \delta_{1,k}\end{bmatrix}.
\]
For each observation $(i,j)$, with $k=S_i^{(m-1)}$, define the row vector
\[
x^{obs}_{ij}=
\begin{bmatrix}
1,\ x_{ij}^\top,\ (1-W_i)D_{ij}^{(m-1)},\ W_iD_{ij}^{(m-1)}x_{ij}^\top,\ W_iD_{ij}^{(m-1)}
\end{bmatrix},
\]
so that $m_{ij,k}(D_{ij}^{(m-1)})=x^{obs}_{ij}B_k+\phi_i^{Y}$.

\paragraph{3(a) Sample $\phi_i^Y$ (conjugate normal).}
Let $k=S_i^{(m-1)}$ and define
$r_{ij}^Y=Y_{ij}^{\mathrm{fact}(m-1)}-x_{ij}^{obs}B_k^{(m-1)}$.
Then
\[
\phi_i^{Y(m)}\mid\cdot \sim N\!\left(M_{\phi_i^Y},V_{\phi_i^Y}\right),
\quad
V_{\phi_i^Y}=\left(\frac{n_i}{\sigma_k^{2(m-1)}}+\frac{1}{\tau_Y^{2(m-1)}}\right)^{-1},
\quad
M_{\phi_i^Y}=V_{\phi_i^Y}\left(\frac{1}{\sigma_k^{2(m-1)}}\sum_{j=1}^{n_i} r_{ij}^Y\right).
\]

\paragraph{3(b) Sample $\tau_Y^2$ (truncated Inv-Gamma).}
Using $\phi_i^Y\sim N(0,\tau_Y^2)$ and $\tau_Y\sim \text{Unif}(0,U_Y)$,
\[
\tau_Y^{2(m)}\mid\cdot \sim \text{Inv-Gamma}\!\left(a_Y,\ b_Y\right)\ \text{truncated to }(0,U_Y^2),
\]
where
\[
a_Y=\frac{I-1}{2},\qquad b_Y=\frac{1}{2}\sum_{i=1}^I \left(\phi_i^{Y(m)}\right)^2.
\]

\paragraph{3(c) Sample $B_k$ (conjugate MVN).}
Stack all $(i,j)$ with $S_i^{(m-1)}=k$.
Let $X_k^{obs}$ be the stacked design matrix and let
\[
\tilde{Y}_k = Y_k^{\mathrm{fact}(m-1)}-\tilde{\phi}_k^{Y(m)},
\]
where $\tilde{\phi}_k^{Y(m)}$ stacks $\phi_i^{Y(m)}\mathbf{1}_{n_i}$ for all $i$ with $S_i^{(m-1)}=k$.
Assume prior $B_k\sim \text{MVN}(0,V_{B,k})$ where $V_{B,k}$ is the block-diagonal covariance implied by the priors on $(\mu_k^Y,\beta_{0,k},\delta_{0,k},\beta_{1,k},\delta_{1,k})$.
Then
\[
B_k^{(m)}\mid\cdot \sim \text{MVN}(M_{Y,k},\Omega_{Y,k}),
\]
with
\[
\Omega_{Y,k}=\left(\frac{1}{\sigma_k^{2(m-1)}}X_k^{obs^\top}X_k^{obs}+V_{B,k}^{-1}\right)^{-1},\qquad
M_{Y,k}=\Omega_{Y,k}\left(\frac{1}{\sigma_k^{2(m-1)}}X_k^{obs^\top}\tilde{Y}_k\right).
\]

\paragraph{3(d) Sample $\sigma_k^2$ (conjugate Inv-Gamma).}
Let $N_k=\sum_{i:S_i^{(m-1)}=k} n_i$ and define residuals
\[
e_{ij,k}=Y_{ij}^{\mathrm{fact}(m-1)}-x^{obs}_{ij}B_k^{(m)}-\phi_i^{Y(m)}.
\]
Then
\[
\sigma_k^{2(m)}\mid\cdot \sim \text{Inv-Gamma}\!\left(a_\sigma+\frac{N_k}{2},\ b_\sigma+\frac{1}{2}\sum_{i:S_i^{(m-1)}=k}\sum_{j=1}^{n_i} e_{ij,k}^2\right).
\]

In the simulation studies, we sample each random-effect
variance by drawing its reciprocal from the corresponding Gamma distribution
with the same shape and rate, truncated below at the reciprocal of the
variance upper bound. We invert the Gamma log survival function for numerical
stability and take the reciprocal of the resulting draw.

\subsubsection*{4. Update latent implementation strata $S_i$}

{
For Steps 4 and 6, let $f_{ij,k}^{(m)}(y;d,w)$ denote the outcome
likelihood with the updated coefficients and outcome random effect,
evaluated under assignment $w$ and compliance type $d$. This is a
Normal density with variance $\sigma_k^{2(m)}$ for continuous outcomes
and a Bernoulli probability with a probit link for binary outcomes.
Let $\eta_{ij,k}^{D(m)}=\mu_k^{D(m)}+x_{ij}^\top\alpha_k^{(m)}
+\phi_i^{D(m)}$. The compliance likelihood integrates out $U_{ij}$;
for binary outcomes the likelihood also integrates out the analogous
outcome augmentation variable. Neither Gaussian augmentation variable
is used again until it is refreshed in the next iteration.
}

For each facility $i$, compute weights for $k=1,\dots,K$:
\begin{align*}
w_{i,k} &\propto \pi_k^{(m)}\,
\text{MVN}\!\left(T_i^{(m)}\mid \mu_k^{S(m)},\Sigma^{(m)}\right)\\
&\quad {}\times
\prod_{j=1}^{n_i}
\left\{\text{Bernoulli}\!\left(D_{ij}^{(m-1)}\mid\Phi(\eta_{ij,k}^{D(m)})\right)
f_{ij,k}^{(m)}(Y_{ij}^{\mathrm{fact}(m-1)};D_{ij}^{(m-1)},W_i)\right\},
\end{align*}
where $T_i^{(m)}=\begin{bmatrix} C_i^{(m)}\\ Z_i\end{bmatrix}$ and $C_i^{(m)}=C_i$ if $W_i=1$ and $C_i^{(m)}=C_i^{mis(m-1)}$ if $W_i=0$.
Normalize $\tilde{w}_{i,k}=w_{i,k}/\sum_{\ell=1}^K w_{i,\ell}$ and sample
\[
S_i^{(m)}\mid\cdot \sim \text{Categorical}(\tilde{w}_{i,1},\dots,\tilde{w}_{i,K}).
\]

\subsubsection*{5. Impute missing facility implementation measures $C_i$ for control facilities}

For each $i$ with $W_i=0$, let $k=S_i^{(m)}$.
Partition $\mu_k^{S(m)}=\begin{bmatrix}\mu_{kC}^{(m)}\\ \mu_{kZ}^{(m)}\end{bmatrix}$ and
\[
\Sigma^{(m)}=
\begin{bmatrix}
\Sigma_{CC}^{(m)} & \Sigma_{CZ}^{(m)}\\
\Sigma_{ZC}^{(m)} & \Sigma_{ZZ}^{(m)}
\end{bmatrix}.
\]
Then the conditional MVN gives
\[
C_i^{mis(m)}\mid Z_i,S_i^{(m)}=k,\mu_k^{S(m)},\Sigma^{(m)}
\sim
\text{MVN}\!\left(M_{C_i},\Omega_{C}\right),
\]
with
\[
M_{C_i}=\mu_{kC}^{(m)}+\Sigma_{CZ}^{(m)}(\Sigma_{ZZ}^{(m)})^{-1}\left(Z_i-\mu_{kZ}^{(m)}\right),
\qquad
\Omega_{C}=\Sigma_{CC}^{(m)}-\Sigma_{CZ}^{(m)}(\Sigma_{ZZ}^{(m)})^{-1}\Sigma_{ZC}^{(m)}.
\]

\subsubsection*{6. Impute unobserved individual compliance $D_{ij}$}

{
For each resident with unobserved $D_{ij}$, let $k=S_i^{(m)}$,
$p_{ij,k}=\Phi(\eta_{ij,k}^{D(m)})$ and
$L_d=f_{ij,k}^{(m)}(Y_{ij}^{\mathrm{fact}(m-1)};d,W_i)$ for $d=0,1$.
The two likelihoods use the resident's actual assignment $W_i$,
including when compliance is missing for a treated resident. Draw
\[
D_{ij}^{mis(m)}\mid\cdot\ \sim\
\mathrm{Bernoulli}\!\left(
\frac{p_{ij,k}L_1}{p_{ij,k}L_1+(1-p_{ij,k})L_0}
\right).
\]
Observed compliance indicators are left unchanged.
}

\subsubsection*{7. Impute missing outcomes under the assigned arm, when present}

For each missing assigned-arm outcome, draw
$Y_{ij}^{\mathrm{fact}(m)}$ from $f_{ij,k}^{(m)}(\,\cdot\,;D_{ij}^{(m)},W_i)$
with $k=S_i^{(m)}$. This step assumes that the missingness mechanism
can be ignored in the likelihood under the stated model. It is not
needed when all assigned-arm outcomes are observed, as in the
simulation studies. Outcomes under the unassigned arm are not
required for these Gibbs updates.

\subsection*{Label switching} \label{app:label_switching}
Finite mixture likelihoods are invariant to permutations of component labels, so MCMC output may exhibit label switching unless labels are constrained during sampling or relabelled after sampling \citep{Stephens2000}. We used a post-processing relabeling approach, leaving the Gibbs sampler itself unchanged. For each retained posterior draw, latent implementation strata were ordered by the first coordinate of the mixture mean vector, so that stratum~1 corresponded to the component with the larger first-coordinate mean and stratum~2 corresponded to the component with the smaller first-coordinate mean. This coordinate separated the two implementation strata in the METRIcAL analyses.

The relabeling rule was applied to all label-dependent quantities before posterior summarization. These included mixture parameters, latent stratum indicators, stratum-specific ITT and CACE estimands, trace plots, convergence diagnostics, and posterior predictive checks within strata. Marginal estimands that average over latent strata are label-invariant, but we applied the same post-processing step before summarizing all retained draws to keep the reported posterior output internally consistent.

\section{Posterior Predictive Imputation of Potential Outcomes}
\label{MissingImputationDerivation}

{
To estimate finite-sample effects, at each posterior draw we draw the
outcome each resident would have had under the other assignment.
We write $\mathbf Y_{\mathrm{cf}}$ for these counterfactual outcomes.
Missing outcomes under the actual assignment are drawn within the
Gibbs sampler instead.

The posterior predictive distribution is
\begin{align}
P(\mathbf Y_{\mathrm{cf}}\mid\mathbf O)
&=\sum_{\mathbf S}\sum_{\mathbf D^{mis}}\int
 P(\mathbf Y_{\mathrm{cf}}\mid\mathbf S,\mathbf D,\mathbf X,
      \mathbf W,\boldsymbol\phi^Y,\boldsymbol\theta)
 \,P(\boldsymbol\theta,\mathbf L\mid\mathbf O)\,
 d\boldsymbol\theta\,d\mathbf L_c,
\label{PosteriorPredictiveInt}
\end{align}
where $\mathbf L_c$ contains the components of $\mathbf L$ other than
$\mathbf S$ and $\mathbf D^{mis}$. The integral over missing binary
outcomes is a sum.

Given the sampled model quantities, covariates, and assignments, we
draw counterfactual outcomes independently across residents. The
Normal draws below are for continuous outcomes. For a binary outcome,
we draw from a Bernoulli distribution with probability $\Phi(\eta)$,
where $\eta$ is the outcome model's linear predictor under the other
assignment.
}

\paragraph{Clusters assigned to control ($W_i=0$)}
For individuals in control clusters, the treated potential outcome is unobserved and is drawn as
\[
Y_{ij}^{(m)}(1,D_{ij}^{(m)}, S_i^{(m)}) \sim
N\!\left(
\mu_{k}^{Y(m)} + \mathbf{X}_{ij}^{\top}\beta_{0,k}^{(m)}
+ D_{ij}^{(m)}\bigl(\mathbf{X}_{ij}^{\top}\beta_{1,k}^{(m)}+\delta_{1,k}^{(m)}\bigr)
+ \phi_{i}^{Y(m)},
\;\sigma_{k}^{2(m)}
\right).
\]

\paragraph{Clusters assigned to treatment ($W_i=1$)}
For individuals in treated clusters, the control potential outcome is unobserved and is drawn as
\[
Y_{ij}^{(m)}(0,0, S_i^{(m)}) \sim
N\!\left(
\mu_{k}^{Y(m)} + \mathbf{X}_{ij}^{\top}\beta_{0,k}^{(m)}
+ D_{ij}^{(m)}\,\delta_{0,k}^{(m)}
+ \phi_{i}^{Y(m)},
\;\sigma_{k}^{2(m)}
\right),
\]
where one-sided noncompliance implies $R_{ij}(0)=0$ (i.e., no receipt under control), but the
control potential outcome is allowed to differ by principal compliance status $D_{ij}$ through
$\delta_{0,k}$.
\paragraph{Posterior predictive procedure}
The posterior predictive imputation proceeds as follows:
\begin{enumerate}
\item For each retained Gibbs draw, obtain
$\boldsymbol\theta^{(m)}$ and $\mathbf L^{(m)}$, including the sampled
facility outcome effects and any imputed assigned-arm outcomes.
\item Conditional on these draws, sample the outcomes under
the unassigned arm according to the distributions above.
\item Compute the finite-sample causal estimands using the quantities outlined in Section \ref{sec:procedures}.
\end{enumerate}

\section{Super-population Estimand Derivation as a Function of Model Parameters}
\label{app:superpopDerivation}

We derive the super-population estimands described in Section~\ref{sec:superpop} as functions of the model parameters under one-sided noncompliance. We use $D_{ij}\in\{0,1\}$ to denote an individual's latent principal stratum type (``complier type''), and we enforce one-sided noncompliance via $R_{ij}(0)=0$ and $R_{ij}(1)=D_{ij}$. Thus, the potential outcomes are indexed as $Y_{ij}(1,D_{ij}, S_i)$ under treatment and $Y_{ij}(0,0, S_i)$ under control, while $Y_{ij}(0,0, S_i)$ may still depend on $D_{ij}$ through principal stratum differences.

{
Let $F_k^\star$ denote the target covariate distribution within
implementation stratum $k$. We consider $F_k^\star=\hat F$ for pooled
standardization and $F_k^\star=\hat F_k$ for stratum-specific
standardization. The expectation $E_{F_k^\star}$ denotes an arithmetic
mean over all $n=\sum_i n_i$ resident rows for $\hat F$, or over the
$\sum_{i\in\mathcal{I}_k}n_i$ resident rows in facilities
$\mathcal{I}_k=\{i:S_i=k\}$ for $\hat F_k$. In the latter case,
$\mathcal{I}_k$ is determined at each posterior draw, and the empirical
distribution is defined only when the stratum contains at least one
resident.

The compliance and outcome random effects are independent under the
model conditional on stratum, covariates, and the model parameters.
The outcome random effect has conditional mean zero and is independent
of compliance given the same quantities. Thus, it averages to zero
in the continuous arm-specific means even among compliers. It also
cancels from the continuous treatment-minus-control contrast because
it is shared across arms. For the binary outcome, we instead integrate
over the outcome random effect on the probability scale.

Integrating over the compliance random effect gives
\[
p_{ijk}
=
\Pr(D_{ij}=1\mid \mathbf{X}_{ij},S_i=k)
=
\Phi\!\left(
\frac{\mathbf{X}_{ij}^{\top}\alpha_k+\mu_k^{D}}{\sqrt{1+\tau_D^2}}
\right),
\qquad
\bar p_k=E_{F_k^\star}[p_{ijk}].
\]
Let $\pi_k$ denote the stratum weights used to define the target
population. If $\pi_k$ denotes the proportion of facilities in stratum $k$,
these weights coincide with the resident proportions when mean
facility size is the same across strata. By Bayes' rule, the stratum
weights among compliers are
\[
\Pr(S_i=k\mid D_{ij}=1)
=
\frac{\pi_k\bar p_k}{\sum_{s=1}^K\pi_s\bar p_s}.
\]
The marginal CACE requires $\sum_{s=1}^K\pi_s\bar p_s>0$, and the
stratum-specific CACE requires $\bar p_k>0$ in the corresponding
stratum.
}

\subsection{Super-population estimands for continuous outcomes}

Write $\delta_k=\delta_{1,k}-\delta_{0,k}$ for the difference between
the two compliance-related outcome coefficients.

\subsubsection{ITT}

\paragraph{$P(X_{ij})=\hat{F}(X_{ij})$}

Under the outcome models,
\[
E\!\left[Y_{ij}(1,D_{ij}, S_i)\mid S_i=k,X_{ij}\right]
=
\mu_k^{Y}+\mathbf{X}_{ij}^{\top}\beta_{0,k}
+
p_{ijk}\bigl(\mathbf{X}_{ij}^{\top}\beta_{1,k}+\delta_{1,k}\bigr),
\]
and
\[
E\!\left[Y_{ij}(0,0,S_i)\mid S_i=k,X_{ij}\right]
=
\mu_k^{Y}+\mathbf{X}_{ij}^{\top}\beta_{0,k}
+
p_{ijk}\delta_{0,k},
\]
where the dependence of $Y_{ij}(0,0,S_i)$ on $D_{ij}$ is through the complier-type contrast $\delta_{0,k}$ and the complier-type probability $p_{ijk}$.

Taking expectations over $S_i\sim\text{Cat}(\pi)$ and $X_{ij}\sim \hat F$, we obtain
\begin{align*}
E[Y_{ij}(1,D_{ij}, S_i)]
&=
\frac{1}{n}\sum_{ij}\sum_{k=1}^{K}\pi_k
\left(
\mu_k^{Y}+\mathbf{X}_{ij}^{\top}\beta_{0,k}
+
p_{ijk}\bigl(\mathbf{X}_{ij}^{\top}\beta_{1,k}+\delta_{1,k}\bigr)
\right),\\
E[Y_{ij}(0,0,S_i)]
&=
\frac{1}{n}\sum_{ij}\sum_{k=1}^{K}\pi_k
\left(
\mu_k^{Y}+\mathbf{X}_{ij}^{\top}\beta_{0,k}
+
p_{ijk}\delta_{0,k}
\right).
\end{align*}
Therefore,
\[
\text{ITT}^{sp}
=
E\!\bigl[Y_{ij}(1,D_{ij}, S_i)-Y_{ij}(0,0,S_i)\bigr]
=
\frac{1}{n}\sum_{ij}\sum_{k=1}^{K}\pi_k\,p_{ijk}\bigl(\mathbf{X}_{ij}^{\top}\beta_{1,k}+\delta_k\bigr).
\]

\paragraph{$P_k(X_{ij})=\hat{F}_k(X_{ij})$}

If $X_{ij}\mid S_i=k \sim \hat F_k$, then
\begin{align*}
E[Y_{ij}(1,D_{ij}, S_i)]
&=
\sum_{k=1}^{K}\pi_k
\left(
\mu_k^{Y}
+
\frac{1}{\sum_{i\in\mathcal{I}_k}n_i}\sum_{i\in\mathcal{I}_k}\sum_{j=1}^{n_i}
\left[
\mathbf{X}_{ij}^{\top}\beta_{0,k}
+
p_{ijk}\bigl(\mathbf{X}_{ij}^{\top}\beta_{1,k}+\delta_{1,k}\bigr)
\right]
\right),\\
E[Y_{ij}(0,0,S_i)]
&=
\sum_{k=1}^{K}\pi_k
\left(
\mu_k^{Y}
+
\frac{1}{\sum_{i\in\mathcal{I}_k}n_i}\sum_{i\in\mathcal{I}_k}\sum_{j=1}^{n_i}
\left[
\mathbf{X}_{ij}^{\top}\beta_{0,k}
+
p_{ijk}\delta_{0,k}
\right]
\right),
\end{align*}
and thus
\[
\text{ITT}^{sp}
=
\sum_{k=1}^{K}\pi_k
\left(
\frac{1}{\sum_{i\in\mathcal{I}_k}n_i}\sum_{i\in\mathcal{I}_k}\sum_{j=1}^{n_i}
p_{ijk}\bigl(\mathbf{X}_{ij}^{\top}\beta_{1,k}+\delta_k\bigr)
\right).
\]

\subsubsection{CACE}

{
Under either covariate distribution, the empirical mean of $\mathbf{X}_{ij}$
among compliers in stratum $k$ is
\[
\overline{\mathbf{X}}_{D,k}
=E[\mathbf{X}_{ij}\mid D_{ij}=1,S_i=k]
=\frac{E_{F_k^\star}[\mathbf{X}_{ij}p_{ijk}]}{\bar p_k}.
\]
For pooled standardization the numerator and denominator average over
all resident rows. For stratum-specific standardization they average
only over residents in facilities $i\in\mathcal{I}_k$. Applying the
complier stratum weights to the conditional outcome means gives
\begin{align*}
E[Y_{ij}(1,D_{ij}, S_i)\mid D_{ij}=1]
&=
\frac{\sum_{k=1}^K\pi_k\bar p_k
\left\{\mu_k^Y+\overline{\mathbf{X}}_{D,k}^{\top}
(\beta_{0,k}+\beta_{1,k})+\delta_{1,k}\right\}}
{\sum_{k=1}^K\pi_k\bar p_k},\\
E[Y_{ij}(0,0,S_i)\mid D_{ij}=1]
&=
\frac{\sum_{k=1}^K\pi_k\bar p_k
\left\{\mu_k^Y+\overline{\mathbf{X}}_{D,k}^{\top}
\beta_{0,k}+\delta_{0,k}\right\}}
{\sum_{k=1}^K\pi_k\bar p_k}.
\end{align*}
Subtracting the arm-specific means and using the definition of
$\overline{\mathbf{X}}_{D,k}$ yields
\[
\text{CACE}^{sp}
=
\frac{\sum_{k=1}^K\pi_k E_{F_k^\star}\!\left[
p_{ijk}\bigl(\mathbf{X}_{ij}^{\top}\beta_{1,k}+\delta_k\bigr)\right]}
{\sum_{k=1}^K\pi_k\bar p_k}
=
\frac{\text{ITT}^{sp}}{\sum_{k=1}^K\pi_k\bar p_k}.
\]
The last equality follows because the continuous outcome model has
the same mean under both assignments for noncompliers, so only
compliers contribute to the ITT numerator.
}

\subsubsection{$\text{ITT}^{sp}_k$}

\paragraph{$P(X_{ij})=\hat{F}(X_{ij})$}

Conditioning on $S_i=k$ and integrating over $X_{ij}\sim \hat F$,
\begin{align*}
E[Y_{ij}(1,D_{ij}, S_i)\mid S_i=k]
&=
\mu_k^{Y}
+
\frac{1}{n}\sum_{ij}\mathbf{X}_{ij}^{\top}\beta_{0,k}
+
\frac{1}{n}\sum_{ij}p_{ijk}\bigl(\mathbf{X}_{ij}^{\top}\beta_{1,k}+\delta_{1,k}\bigr),\\
E[Y_{ij}(0,0,S_i)\mid S_i=k]
&=
\mu_k^{Y}
+
\frac{1}{n}\sum_{ij}\mathbf{X}_{ij}^{\top}\beta_{0,k}
+
\frac{1}{n}\sum_{ij}p_{ijk}\delta_{0,k}.
\end{align*}
Thus,
\[
\text{ITT}^{sp}_k
=
E[Y_{ij}(1,D_{ij}, S_i)-Y_{ij}(0,0,S_i)\mid S_i=k]
=
\frac{1}{n}\sum_{ij}p_{ijk}\bigl(\mathbf{X}_{ij}^{\top}\beta_{1,k}+\delta_k\bigr).
\]

\paragraph{$P_k(X_{ij})=\hat{F}_k(X_{ij})$}

If $X_{ij}\mid S_i=k\sim \hat F_k$, then
\begin{align*}
E[Y_{ij}(1,D_{ij}, S_i)\mid S_i=k]
&=
\mu_k^{Y}
+
\frac{1}{\sum_{i\in\mathcal{I}_k}n_i}\sum_{i\in\mathcal{I}_k}\sum_{j=1}^{n_i}
\left[
\mathbf{X}_{ij}^{\top}\beta_{0,k}
+
p_{ijk}\bigl(\mathbf{X}_{ij}^{\top}\beta_{1,k}+\delta_{1,k}\bigr)
\right],\\
E[Y_{ij}(0,0,S_i)\mid S_i=k]
&=
\mu_k^{Y}
+
\frac{1}{\sum_{i\in\mathcal{I}_k}n_i}\sum_{i\in\mathcal{I}_k}\sum_{j=1}^{n_i}
\left[
\mathbf{X}_{ij}^{\top}\beta_{0,k}
+
p_{ijk}\delta_{0,k}
\right],
\end{align*}
and therefore
\[
\text{ITT}^{sp}_k
=
\frac{1}{\sum_{i\in\mathcal{I}_k}n_i}\sum_{i\in\mathcal{I}_k}\sum_{j=1}^{n_i}
p_{ijk}\bigl(\mathbf{X}_{ij}^{\top}\beta_{1,k}+\delta_k\bigr).
\]

\subsubsection{$\text{CACE}^{sp}_k$}

\paragraph{$P(X_{ij})=\hat{F}(X_{ij})$}

For a fixed type $k$,
\begin{align*}
E[Y_{ij}(1,D_{ij}, S_i)\mid D_{ij}=1,S_i=k]
&=
\mu_k^{Y}
+
\frac{\frac{1}{n}\sum_{ij}\mathbf{X}_{ij}^{\top}p_{ijk}}{\frac{1}{n}\sum_{ij}p_{ijk}}(\beta_{0,k}+\beta_{1,k})
+
\delta_{1,k},\\
E[Y_{ij}(0,0,S_i)\mid D_{ij}=1,S_i=k]
&=
\mu_k^{Y}
+
\frac{\frac{1}{n}\sum_{ij}\mathbf{X}_{ij}^{\top}p_{ijk}}{\frac{1}{n}\sum_{ij}p_{ijk}}\beta_{0,k}
+
\delta_{0,k}.
\end{align*}
Thus,
\[
\text{CACE}^{sp}_k
=
E[Y_{ij}(1,D_{ij}, S_i)-Y_{ij}(0,0,S_i)\mid D_{ij}=1,S_i=k]
=
\frac{\frac{1}{n}\sum_{ij}\mathbf{X}_{ij}^{\top}\beta_{1,k}p_{ijk}}{\frac{1}{n}\sum_{ij}p_{ijk}}
+
\delta_k.
\]

\paragraph{$P_k(X_{ij})=\hat{F}_k(X_{ij})$}

If $X_{ij}\mid S_i=k\sim \hat F_k$, then
\begin{align*}
E[Y_{ij}(1,D_{ij}, S_i)\mid D_{ij}=1,S_i=k]
&=
\mu_k^{Y}
+
\frac{
\frac{1}{\sum_{i\in\mathcal{I}_k}n_i}\sum_{i\in\mathcal{I}_k}\sum_{j=1}^{n_i}\mathbf{X}_{ij}^{\top}p_{ijk}
}{
\frac{1}{\sum_{i\in\mathcal{I}_k}n_i}\sum_{i\in\mathcal{I}_k}\sum_{j=1}^{n_i}p_{ijk}
}(\beta_{0,k}+\beta_{1,k})
+
\delta_{1,k},\\
E[Y_{ij}(0,0,S_i)\mid D_{ij}=1,S_i=k]
&=
\mu_k^{Y}
+
\frac{
\frac{1}{\sum_{i\in\mathcal{I}_k}n_i}\sum_{i\in\mathcal{I}_k}\sum_{j=1}^{n_i}\mathbf{X}_{ij}^{\top}p_{ijk}
}{
\frac{1}{\sum_{i\in\mathcal{I}_k}n_i}\sum_{i\in\mathcal{I}_k}\sum_{j=1}^{n_i}p_{ijk}
}\beta_{0,k}
+
\delta_{0,k}.
\end{align*}
Therefore,
\[
\text{CACE}^{sp}_k
=
E[Y_{ij}(1,D_{ij}, S_i)-Y_{ij}(0,0,S_i)\mid D_{ij}=1,S_i=k]
=
\frac{
\frac{1}{\sum_{i\in\mathcal{I}_k}n_i}\sum_{i\in\mathcal{I}_k}\sum_{j=1}^{n_i}\mathbf{X}_{ij}^{\top}\beta_{1,k}p_{ijk}
}{
\frac{1}{\sum_{i\in\mathcal{I}_k}n_i}\sum_{i\in\mathcal{I}_k}\sum_{j=1}^{n_i}p_{ijk}
}
+
\delta_k.
\]

\subsubsection{Summary}

Table~\ref{EstimandFuncs} summarizes the continuous-outcome derivations.

\begin{table}[H]
{\centering

\renewcommand{\arraystretch}{1.25}
\setlength{\tabcolsep}{6pt}

\begin{tabular}{|c|c|c|}
\hline
\multirow{2}{*}{Estimand} 
& $P_k(\mathbf{X}_{ij})=\hat{F}(\mathbf{X}_{ij})$ 
& $P_k(\mathbf{X}_{ij})=\hat{F}_k(\mathbf{X}_{ij})$ \\
& (pooled super-population) 
& (stratum-specific super-population) \\
\hline
\hline
$\text{ITT}^{sp}$ 
& $\displaystyle
\frac{1}{n}\sum_{ij}\sum_k \pi_k\, p_{ijk}
\left(\mathbf{X}_{ij}^{\top}\beta_{1,k}+\delta_k\right)$
& $\displaystyle
\sum_k \pi_k
\left(
\frac{1}{\sum_{i\in\mathcal{I}_k} n_i}
\sum_{i\in\mathcal{I}_k}\sum_{j=1}^{n_i}
p_{ijk}\left(\mathbf{X}_{ij}^{\top}\beta_{1,k}+\delta_k\right)
\right)$ \\
\hline
$\text{ITT}^{sp}_k$ 
& $\displaystyle
\frac{1}{n}\sum_{ij}
p_{ijk}\left(\mathbf{X}_{ij}^{\top}\beta_{1,k}+\delta_k\right)$
& $\displaystyle
\frac{1}{\sum_{i\in\mathcal{I}_k} n_i}
\sum_{i\in\mathcal{I}_k}\sum_{j=1}^{n_i}
p_{ijk}\left(\mathbf{X}_{ij}^{\top}\beta_{1,k}+\delta_k\right)$ \\
\hline
$\text{CACE}^{sp}$
& $\displaystyle
\frac{\text{ITT}^{sp}}{\sum_k\pi_k\bar p_k}$
& $\displaystyle
\frac{\text{ITT}^{sp}}{\sum_k\pi_k\bar p_k}$ \\
\hline
$\text{CACE}^{sp}_k$ 
& $\displaystyle
\frac{
\sum_{ij} p_{ijk}
\left(\mathbf{X}_{ij}^{\top}\beta_{1,k}+\delta_k\right)
}{
\sum_{ij} p_{ijk}
}$
& $\displaystyle
\frac{
\sum_{i\in\mathcal{I}_k}\sum_{j=1}^{n_i}
\mathbf{X}_{ij}^{\top}\beta_{1,k} p_{ijk}
}{
\sum_{i\in\mathcal{I}_k}\sum_{j=1}^{n_i}
p_{ijk}
}
+ \delta_k$ \\
\hline
\end{tabular}

\begin{minipage}{16.5cm}
\vspace{6pt}
\footnotesize
$p_{ijk}=\Pr(D_{ij}=1\mid \mathbf{X}_{ij},S_i=k)
=\text{probit}^{-1}\!\left(
(\mathbf{X}_{ij}^{\top}\alpha_k+\mu_k^D)/\sqrt{\tau_D^2+1}
\right)$; \\
$\delta_k=\delta_{1,k}-\delta_{0,k}$;
$\bar p_k=E_{F_k^\star}[p_{ijk}]$ uses the covariate distribution
in the corresponding column; 
$\mathcal{I}_k=\{i:S_i=k\}$. \\
Under one-sided noncompliance, $R_{ij}(0)=0$ for all units, so control potential
outcomes are indexed as $Y_{ij}(0,0,S_i)$; dependence on $D_{ij}$ arises through
principal stratum membership.
\end{minipage}

\belowtablecaption{Super-population estimands for continuous outcomes expressed as functions of the model parameters under alternative choices of the super-population covariate distribution.}
\label{EstimandFuncs}
\par}
\end{table}

\subsection{Super-population estimands for binary outcomes}
\label{app:binarySuperpopDerivation}

{
For the binary probit outcome model, the super-population estimands
are differences in outcome probabilities. Let $\eta_{w,ijk}(d)$
denote the outcome linear predictor under assignment $w$ for a resident
with complier type $D_{ij}=d$ in latent implementation stratum $k$,
excluding the cluster-specific random effect $\phi_i^Y$. For the
antipsychotic outcome model,
\begin{align*}
\eta_{1,ijk}(d)
&= \mu_k^Y + \mathbf{X}_{ij}^{\top}\beta_0
   + d\bigl(\mathbf{X}_{ij}^{\top}\beta_1+\delta_{1,k}\bigr),\\
\eta_{0,ijk}(d)
&= \mu_k^Y + \mathbf{X}_{ij}^{\top}\beta_0 + d\delta_{0,k}.
\end{align*}
Integrating over
$\phi_i^Y\sim \mathcal{N}(0,\tau_Y^2)$ gives the outcome probability
\[
q_{w,ijk}(d)
=
\Phi\!\left(\frac{\eta_{w,ijk}(d)}{\sqrt{1+\tau_Y^2}}\right).
\]
Under the specified outcome model, $q_{1,ijk}(0)=q_{0,ijk}(0)$.

Using the empirical average $E_{F_k^\star}$ defined above, write
\[
A_k=E_{F_k^\star}\!\left[
p_{ijk}\{q_{1,ijk}(1)-q_{0,ijk}(1)\}\right].
\]
Since the outcome probabilities are equal across arms for
noncompliers, the ITT and CACE estimands are
\begin{align*}
\text{ITT}^{sp}_k &= A_k,
&\qquad \text{ITT}^{sp} &= \sum_{k=1}^K \pi_k A_k,\\
\text{CACE}^{sp}_k &= \frac{A_k}{\bar p_k},
&\qquad \text{CACE}^{sp} &=
\frac{\sum_{k=1}^K \pi_k A_k}{\sum_{k=1}^K \pi_k \bar p_k},
\end{align*}
where the CACE expressions require positive complier probabilities
in the corresponding target populations.

Within stratum $k$, the arm-specific outcome probabilities among all
residents and among compliers, respectively, are
\[
E_{F_k^\star}\!\left[p_{ijk}q_{w,ijk}(1)
  +(1-p_{ijk})q_{w,ijk}(0)\right],
\qquad
\frac{E_{F_k^\star}\!\left[p_{ijk}q_{w,ijk}(1)\right]}{\bar p_k}.
\]
The corresponding marginal probabilities average these
stratum-specific probabilities with weights $\pi_k$ among all
residents and weights $\pi_k\bar p_k/\sum_{s=1}^K\pi_s\bar p_s$
among compliers.

These expressions can be evaluated directly at each posterior draw.
}

\section{Simulation Case Studies Details} \label{app:simulation_study}

This appendix section describes the data-generating mechanisms (DGMs) and prior distributions used for modeling in the simulated case studies.

\subsection{Data-Generating Mechanism}

Across all case studies, data are generated under a two-stage structure with latent
implementation strata, individual-level noncompliance, and cluster-correlated outcomes.
Case Studies~2--5 vary model components or latent-stratum separation relative to Case Study~1. Case Study~6 varies cluster sizes while retaining the Case Study~1 data-generating mechanism.

\subsubsection{Case Study 1: Correctly Specified Data-Generating Mechanism}

\paragraph{Latent implementation strata and cluster-level variables.}
Clusters are assigned to one of two latent implementation strata with equal probability,
\begin{align}
S_i &\sim \text{Categorical}(K=2,\; \pi=[0.5,0.5]).
\end{align}
Conditional on $S_i$, cluster-level implementation metrics $\mathbf{C}_i$ and baseline characteristics $Z_i$
are generated from a bivariate normal mixture,
\begin{align}
\begin{bmatrix}
C_i \\
Z_i
\end{bmatrix}
\;|\; S_i=k
&\sim \text{MVN}(\mathbf{m}_k, V), \\
\mathbf{m}_1 &= (-2,-2)^\top, \qquad
\mathbf{m}_2 = (2,2)^\top,
\end{align}
with covariance matrix
\begin{align}
V &=
\begin{bmatrix}
v_C^2 & \rho v_C v_Z \\
\rho v_C v_Z & v_Z^2
\end{bmatrix},
\qquad
v_C^2,v_Z^2 \sim \text{Unif}(0.5,2), \quad
\rho \sim \text{Unif}(-0.8,0.8).
\end{align}

\paragraph{Individual-level compliance.}
Under one-sided noncompliance, individual principal compliance types are defined as
$D_{ij} = R_{ij}(1)$.
Conditional on latent implementation strata, individual covariates, and a cluster-specific effect,
compliance is generated as
\begin{align}
D_{ij} \mid S_i=k, \mathbf{X}_{ij}, \varphi_i^D
&\sim \text{Bernoulli}\!\left(
\Phi\!\left(m_k^D + \mathbf{X}_{ij}^\top \gamma_k + \varphi_i^D\right)
\right), \\
m_1^D &= 0, \quad m_2^D = 0.5, \\
\gamma_1 &= (-0.25,-0.25)^\top, \quad
\gamma_2 = (-0.5,-0.5)^\top, \\
\varphi_i^D &\sim \mathcal{N}(0,\nu_D^2), \qquad \nu_D^2 = 0.25.
\end{align}

\paragraph{Potential outcomes.}
Potential outcomes depend on latent implementation strata, individual covariates, compliance status,
and a cluster-specific outcome effect. For clusters with $S_i=k$,
\begin{align}
Y_{ij}(1,D_{ij}, S_i) \mid S_i=k, D_{ij}, \mathbf{X}_{ij}, \varphi_i^Y
&\sim \mathcal{N}\!\left(
m_k^Y + \mathbf{X}_{ij}^\top \eta_{0,k}
+ D_{ij}(\mathbf{X}_{ij}^\top \eta_{1,k} + \Delta_{1,k})
+ \varphi_i^Y,\;
v_k^2
\right), \\
Y_{ij}(0,0, S_i) \mid S_i=k, D_{ij}, \mathbf{X}_{ij}, \varphi_i^Y
&\sim \mathcal{N}\!\left(
m_k^Y + \mathbf{X}_{ij}^\top \eta_{0,k}
+ D_{ij}\Delta_{0,k}
+ \varphi_i^Y,\;
v_k^2
\right),
\end{align}
with parameters
\begin{align}
m_1^Y &= 2, \quad m_2^Y = 4, \\
v_1^2 &= v_2^2 = 16, \\
\Delta_{0,1} &= 1, \quad \Delta_{0,2} = 2, \\
\Delta_{1,1} &= 5.5, \quad \Delta_{1,2} = 7.5, \\
\eta_{0,1} &= (1,1)^\top, \quad \eta_{0,2} = (2,2)^\top, \\
\eta_{1,1} &= (1,1)^\top, \quad \eta_{1,2} = (2,2)^\top, \\
\varphi_i^Y &\sim \mathcal{N}(0,\nu_Y^2), \qquad \nu_Y^2 = 9.
\end{align}

The influence of baseline covariates and compliance on outcomes differs across latent
implementation strata.

\subsubsection{Case Study 2: Misspecified Cluster-Level Mixture Model}

To assess sensitivity to misspecification of the cluster-level mixture distribution,
we generate $(C_i,Z_i)$ from a bivariate skew-$t$ distribution with 5 degrees of freedom
and slant vector $(2,2)^\top$:
\begin{align}
\begin{bmatrix}
C_i \\
Z_i
\end{bmatrix}
\;|\; S_i=k
&\sim \text{Skew-}t_5\!\left(\mathbf{m}_k,V,(2,2)^\top\right).
\end{align}
We use the Case Study~1 mean vector $\mathbf{m}_k$ and covariance matrix $V$
as the skew-$t$ location and dispersion parameters, respectively.
Individual-level compliance and potential outcomes are generated according to the correctly
specified models in Case Study~1.

\subsubsection{Case Study 3: Misspecified Individual Compliance Model}

Sensitivity to misspecification of the individual compliance model is evaluated by replacing
the probit link with a Burr link function,
\[
\text{Burr}_c(x) = 1 - (1+e^x)^{-c},
\]
with $c=0.5$, yielding compliance probabilities skewed toward zero.
Cluster-level variables and potential outcomes are generated as in Case Study~1.

\subsubsection{Case Study 4: Misspecified Outcome Model}

Outcome model misspecification is introduced by adding interaction terms between individual
baseline covariates.
Conditional on $S_i=k$, $D_{ij}$, $\mathbf{X}_{ij}$, and $\varphi_i^Y$,
\begin{align}
Y_{ij}(1,D_{ij}, S_i)
&\sim \mathcal{N}\!\Bigg(
\begin{aligned}[t]
m_k^Y + \mathbf{X}_{ij}^\top \eta_{0,k}
&-2(X_{ij1}X_{ij2})\\
&+ D_{ij}\big[\mathbf{X}_{ij}^\top \eta_{1,k} + \Delta_{1,k}
-2(X_{ij1}X_{ij2})\big]
+ \varphi_i^Y,
\end{aligned}\;
v_k^2
\Bigg), \\
Y_{ij}(0,0,S_i)
&\sim \mathcal{N}\!\Big(
m_k^Y + \mathbf{X}_{ij}^\top \eta_{0,k}
-2(X_{ij1}X_{ij2})
+ D_{ij}\Delta_{0,k}
+ \varphi_i^Y,\;
v_k^2
\Big).
\end{align}

The interaction coefficients are set to $-2$ in both strata.
Cluster-level variables and compliance are generated as in Case Study~1.

\subsubsection{Case Study 5: Weak Latent-Stratum Separation}

Case Study~5 follows the correctly specified data-generating mechanism in Case Study~1, except that the latent-stratum-specific means for the cluster-level implementation measure and baseline covariate are closer. Specifically, the centered component means for $(C_i, Z_i)$ are $(-0.75,-0.75)^\top$ and $(0.75,0.75)^\top$, rather than $(-2,-2)^\top$ and $(2,2)^\top$ as in Case Study~1. All other components of the data-generating mechanism are unchanged from Case Study~1.

\subsubsection{Case Study 6: Unequal Cluster Sizes}

Case Study~6 follows the correctly specified data-generating mechanism in Case Study~1, except that each treatment arm contains 15 clusters of size 10 and 15 clusters of size 30. Cluster sizes are fixed independently of latent implementation strata, and the total sample size remains 1{,}200. Table~\ref{tab:case6_results} reports operating characteristics across 2{,}500 replications.

\begin{table}[H]
\centering
{
\renewcommand{\arraystretch}{0.75}
\setlength{\tabcolsep}{6pt}
\begin{tabular}{lcccccc}
\hline
Estimand & Coverage (\%) & Bias & StdBias & IW & RMSE & Med. SD Ratio \\
\hline
$\text{ITT}^{sp}$ & 96 & 0.20 & 0.08 & 2.42 & 0.61 & 1.05 \\
$\text{ITT}^{sp}_{1}$ & 97 & 0.17 & 0.08 & 3.22 & 0.76 & 1.10 \\
$\text{ITT}^{sp}_{2}$ & 95 & 0.22 & 0.07 & 3.46 & 0.89 & 1.00 \\
$\text{CACE}^{sp}$ & 96 & 0.36 & 0.08 & 3.96 & 1.01 & 1.05 \\
$\text{CACE}^{sp}_{1}$ & 97 & 0.35 & 0.09 & 6.08 & 1.43 & 1.10 \\
$\text{CACE}^{sp}_{2}$ & 96 & 0.35 & 0.08 & 5.02 & 1.31 & 0.99 \\
\hline
\end{tabular}

\begin{minipage}{12.2cm}
\footnotesize
Coverage: empirical coverage of the 95\% posterior credible interval.
StdBias: bias divided by the absolute value of the true estimand.
IW: mean 95\% credible interval width.
RMSE: root mean squared error.
Med. SD Ratio: median posterior standard deviation across simulation replications divided by the empirical standard deviation of the posterior means across simulation replications.
\end{minipage}
}
\belowtablecaption{Operating characteristics under correct model specification with unequal cluster sizes (Case Study 6) across 2{,}500 replications.}
\label{tab:case6_results}
\end{table}

\subsection{Prior distributions for simulation studies}

This section specifies the prior distributions used in the simulated case studies
described in Section~\ref{sec:CaseStudies}.
The prior distributions are proper and held fixed across
simulation scenarios.

\subsubsection{Latent implementation strata}

Latent implementation strata are modeled as
\[
S_i \sim \mathrm{Categorical}(\pi_1,\pi_2),
\qquad
(\pi_1,\pi_2) \sim \mathrm{Dirichlet}(3,3).
\]

Conditional on $S_i=k$, cluster-level baseline covariates and implementation measures
are modeled as
\[
\begin{pmatrix}
Z_i \\ C_i
\end{pmatrix}
\mid S_i = k
\sim \mathrm{MVN}(\boldsymbol{\mu}^{S}_{k}, \Sigma),
\qquad k=1,2,
\]
with priors
\[
\boldsymbol{\mu}^{S}_{k}
\sim \mathrm{MVN}\!\left(
\begin{pmatrix}0\\0\end{pmatrix},
100 \cdot \mathbf{I}_{2}
\right),
\qquad
\Sigma \sim \mathrm{Inv\text{-}Wishart}\!\left(
\Psi=\tfrac{1}{100}\mathbf{I}_{2},\;
\nu_\Sigma = 7
\right).
\]

\subsubsection{Individual compliance model}

Individual compliance indicators are modeled as
\[
D_{ij} \mid S_i=k, X_{ij}, \phi_i^{D}
\sim \mathrm{Bernoulli}\!\left(
\mathrm{probit}^{-1}\!\left(
\mu_k^{D} + X_{ij}^\top \alpha_k + \phi_i^{D}
\right)\right).
\]

We assign priors
\[
\mu_k^{D} \sim \mathcal{N}(0,100),
\qquad
\alpha_k \sim \mathrm{MVN}(0,100\,\mathbf{I}_{p}),
\]
\[
\phi_{i}^{D} \sim \mathcal{N}(0,\tau_{D}^2),
\qquad
\tau_{D} \sim \mathrm{Uniform}(0,\sqrt{10}).
\]

\subsection{Outcome model}

For the outcome models, we assign priors
\[
\mu_k^{Y} \sim \mathcal{N}(0,100),
\qquad
\delta_{0,k} \sim \mathcal{N}(0,100),
\qquad
\delta_{1,k} \sim \mathcal{N}(0,100),
\]
\[
\beta_{0,k} \sim \mathrm{MVN}(0,100\,\mathbf{I}_{p}),
\qquad
\beta_{1,k} \sim \mathrm{MVN}(0,100\,\mathbf{I}_{p}),
\]
\[
\phi_i^{Y} \sim \mathcal{N}(0,\tau_{Y}^2),
\qquad
\tau_{Y} \sim \mathrm{Uniform}(0,15),
\qquad
\sigma_k^2 \sim \mathrm{Inv\text{-}Gamma}(1,1),
\qquad k=1,2.
\]

The Dirichlet prior favors balanced stratum proportions, and
the inverse-Wishart prior places most of its mass on small within-stratum
variances. The simulation results describe performance under these prior choices.

\section{Prior distributions for METRIcAL application}
\label{app:priors_application}

This section documents the prior distributions used in the analysis of the
METRIcAL trial described in Section~\ref{DataAnalysis}.
Compared to the simulation study priors (Appendix~\ref{app:simulation_study}), these priors are moderately
more informative to improve stability and identifiability in the observed data
analysis while remaining weakly informative relative to the scale of the outcomes.

\subsection{Latent implementation model}

Latent implementation strata are modeled as
\[
S_i \sim \mathrm{Categorical}(\pi_1,\pi_2),
\qquad
(\pi_1,\pi_2) \sim \mathrm{Diriclet}(1,1).
\]

Conditional on $S_i=k$, the facility-level implementation measure (logit clinical
proportion) and baseline facility characteristic (average daily census) follow
\[
\begin{pmatrix}
C_i \\ Z_i
\end{pmatrix}
\mid S_i=k
\sim \mathrm{MVN}(\boldsymbol{\mu}^S_k,\Sigma),
\qquad k=1,2,
\]
with priors
\[
\boldsymbol{\mu}^S_k \sim
\mathrm{MVN}\!\left(
\begin{pmatrix}0\\0\end{pmatrix},
\begin{pmatrix}
1000 & 0\\
0 & 10000
\end{pmatrix}
\right),
\]
\[
\Sigma \sim \mathrm{Inv\text{-}Wishart}\!\left(
\begin{pmatrix}
1/1000 & 0\\
0 & 1/1000
\end{pmatrix},
\;5
\right).
\]

This specification allows implementation intensity and facility size to be correlated
within latent implementation strata while sharing a common covariance structure.

\subsection{Individual compliance model}

Individual-level compliance is modeled using a multilevel probit regression:
\[
D_{ij} \mid S_i=k, X_{ij}, \phi_i^{D}
\sim
\mathrm{Bernoulli}\!\left(
\mathrm{probit}^{-1}\!\left(
\mu_k^D + X_{ij}^\top \alpha + \phi_i^D
\right)\right),
\]
with priors
\[
\mu_k^D \sim \mathcal{N}(0,3^2),
\qquad
\alpha \sim \mathrm{MVN}(0,10\,\mathbf{I}),
\]
\[
\phi_i^D \sim \mathcal{N}(0,\tau_D^2),
\qquad
\tau_D \sim \mathrm{Uniform}(0,4).
\]

The intercept $\mu_k^D$ is allowed to vary across latent implementation strata, while
covariate effects and residual cluster-level variability are shared across types to
improve identifiability.

\subsection{CMAI outcome model}

Potential outcomes for change in CMAI are modeled as
\begin{align*}
Y_{ij}(1,D_{ij}, S_i) \mid S_i=k, D_{ij}, X_{ij}, \phi_i^{Y} &\sim
\mathcal{N}\!\left(
\mu_k^Y + X_{ij}^\top \beta_0
+ D_{ij}\bigl(X_{ij}^\top \beta_1 + \delta_{1,k}\bigr)
+ \phi_i^Y,
\;\sigma^2
\right),\\
Y_{ij}(0,0, S_i) \mid S_i=k, D_{ij}, X_{ij}, \phi_i^{Y} &\sim
\mathcal{N}\!\left(
\mu_k^Y + X_{ij}^\top \beta_0
+ D_{ij}\delta_{0,k}
+ \phi_i^Y,
\;\sigma^2
\right),
\end{align*}
with priors
\[
\mu_k^Y \sim \mathcal{N}(0,100^2),
\qquad
\delta_{1,k} \sim \mathcal{N}(0,10^2),
\qquad
\delta_{0,k} \sim \mathcal{N}(0,4^2),
\]
\[
\beta_0 \sim \mathrm{MVN}(0,10^2\,\mathbf{I}),
\qquad
\beta_1 \sim \mathrm{MVN}(0,10^2\,\mathbf{I}),
\]
\[
\phi_i^Y \sim \mathcal{N}(0,\tau_Y^2),
\qquad
\tau_Y \sim \mathrm{Uniform}(0,15),
\qquad
\sigma^2 \sim \mathrm{Inv\text{-}Gamma}(1,1).
\]

Intercepts and compliance-related contrasts are allowed to vary across latent
implementation strata, while covariate effects and variance components are shared.

\subsection{Antipsychotic outcome model}

Antipsychotic use is modeled using a probit regression:
\begin{align*}
Y_{ij}(1,D_{ij}, S_i) \mid S_i=k, D_{ij}, X_{ij}, \phi_i^{Y} &\sim
\mathrm{Bernoulli}\!\left(
\mathrm{probit}^{-1}\!\left(
\mu_k^Y + X_{ij}^\top \beta_0
+ D_{ij}\bigl(X_{ij}^\top \beta_1 + \delta_{1,k}\bigr)
+ \phi_i^Y
\right)\right),\\
Y_{ij}(0,0,S_i) \mid S_i=k, D_{ij}, X_{ij}, \phi_i^{Y} &\sim
\mathrm{Bernoulli}\!\left(
\mathrm{probit}^{-1}\!\left(
\mu_k^Y + X_{ij}^\top \beta_0
+ D_{ij}\delta_{0,k}
+ \phi_i^Y
\right)\right),
\end{align*}
with priors
\[
\mu_k^Y \sim \mathcal{N}(0,5^2),
\qquad
\delta_{1,k} \sim \mathcal{N}(0,5^2),
\qquad
\delta_{0,k} \sim \mathcal{N}(0,1^2),
\]
\[
\beta_0 \sim \mathrm{MVN}(0,5^2\,\mathbf{I}),
\qquad
\beta_1 \sim \mathrm{MVN}(0,5^2\,\mathbf{I}),
\]
\[
\phi_i^Y \sim \mathcal{N}(0,\tau_Y^2),
\qquad
\tau_Y \sim \mathrm{Uniform}(0,\sqrt{10}).
\]

The tighter prior on $\delta_{0,k}$ reflects the assumption that, conditional on
baseline covariates, differences between compliers and noncompliers under standard
care are unlikely to be large on the probit scale.

\section{Supplementary Material for Application} \label{app:Application}
We include supplementary material for the application section below.

\begin{table}[h]
{\centering
{
\renewcommand{\arraystretch}{0.85}
\begin{tabular}{lccc}
\hline
Baseline characteristic & Total & Treatment & Control \\
 & $N=976$ & $N_1=483$ & $N_0=493$ \\
\hline
Age, y, mean (SD) & 80.3 (12.3) & 79.8 (12.2) & 80.8 (12.3) \\
Female sex, $n$ (\%) & 676 (69.3) & 327 (67.7) & 349 (70.8) \\
White race, $n$ (\%) & 708 (72.8) & 345 (71.7) & 363 (73.9) \\
Severe ADL impairment (score $\geq 21$), $n$ (\%) & 281 (28.8) & 162 (33.5) & 119 (24.2) \\
Cognitive function, $n$ (\%) & & & \\
\quad Cognitively intact & 113 (11.6) & 52 (10.8) & 61 (12.4) \\
\quad Mildly impaired & 206 (21.1) & 102 (21.1) & 104 (21.1) \\
\quad Moderately impaired & 509 (52.1) & 252 (52.2) & 257 (52.1) \\
\quad Severely impaired & 148 (15.2) & 77 (15.9) & 71 (14.4) \\
Depression, $n$ (\%) & 559 (57.3) & 264 (54.7) & 295 (59.8) \\
Any antipsychotic use in the past week, $n$ (\%) & 289 (29.6) & 122 (25.3) & 167 (33.9) \\
\hline
\end{tabular}

\begin{minipage}{12cm}
\footnotesize
Values are from Table~1 of \citet{McCreedy2022}. Race was missing for two
residents in each arm, and ADL was missing for one control resident.
Percentages for those characteristics use nonmissing denominators.
ADL: Activities of Daily Living.
\end{minipage}
}
\par}

\belowtablecaption{Mean and standard deviation of resident-level characteristics derived from the MDS}
\label{tab:XSummary}
\end{table}

\begin{table}[h]
{\centering
    
\renewcommand{\arraystretch}{0.75}
\begin{tabular}{ccccc}
\hline 
\multirow{2}{*}{Metric/Characteristic} & \multicolumn{2}{c}{Intervention} & \multicolumn{2}{c}{Control}\tabularnewline
\cline{2-5} 
 & Mean & SD & Mean & SD\tabularnewline
\hline 
\multicolumn{1}{r}{Clinical Relevance$^{\dagger}$} & 46.8 & 26.5 & - & -\tabularnewline
\multicolumn{1}{r}{$\ $Average Daily Census} & 62.0 & 24.5 & 63.6 & 25.6\tabularnewline
\hline 

\end{tabular}
\\
\begin{minipage}{10cm}\linespread{0.8}
\footnotesize
$^\dagger$ These percentages are modeled on the logit scale, and transformed back to percentages for interpretation
\end{minipage}
\par}

\linespread{1.0}
     \belowtablecaption{Mean and standard deviation of facility characteristics and facility compliance metrics used in each model pair}
 \label{tab:CZSummary}
\end{table}

MCMC diagnostics for each estimand are shown in Table~\ref{GRstats}.

\begin{table}[H] 

{\centering
    
\renewcommand{\arraystretch}{0.85}
\setlength{\tabcolsep}{5pt}
\begin{tabular}{lccc} 
\hline 
Outcome & Estimand & $\hat{R}$ & ESS\tabularnewline
\hline 
\multicolumn{4}{l}{CMAI}\tabularnewline
& ITT & 1.002 & 1144\tabularnewline
& $\text{ITT}_{1}$ & 1.000 & 1892\tabularnewline
& $\text{ITT}_{2}$ & 1.003 & 874\tabularnewline
& $\text{ITT}_{1}-\text{ITT}_{2}$ & 1.001 & 1255\tabularnewline
& CACE & 1.001 & 1606\tabularnewline
& $\text{CACE}_{1}$ & 1.000 & 2219\tabularnewline
& $\text{CACE}_{2}$ & 1.002 & 1007\tabularnewline
& $\text{CACE}_{1}-\text{CACE}_{2}$ & 1.001 & 1593\tabularnewline
\hline 
\multicolumn{4}{l}{Antipsychotics}\tabularnewline
& ITT & 1.004 & 466\tabularnewline
& $\text{ITT}_{1}$ & 1.009 & 325\tabularnewline
& $\text{ITT}_{2}$ & 1.003 & 434\tabularnewline
& $\text{ITT}_{1}-\text{ITT}_{2}$ & 1.006 & 365\tabularnewline
& CACE & 1.003 & 412\tabularnewline
& $\text{CACE}_{1}$ & 1.014 & 270\tabularnewline
& $\text{CACE}_{2}$ & 1.004 & 397\tabularnewline
& $\text{CACE}_{1}-\text{CACE}_{2}$ & 1.011 & 274\tabularnewline
\hline 
\end{tabular}
\par}

\linespread{1.0}
     \belowtablecaption{MCMC diagnostics for the 4 chains of posterior samples for each estimand presented in the data analysis in Section~\ref{DataAnalysis}. $\hat{R}$ denotes the Gelman--Rubin potential scale reduction factor, and ESS denotes the effective sample size across retained posterior draws.}
\label{GRstats}
\end{table}
For each outcome analysis, we ran 4 MCMC chains for 3{,}500 iterations per chain, discarded the first 500 iterations as burn-in, and retained every third subsequent draw to obtain 1{,}000 posterior draws per chain. Relabeling stability was assessed by checking that the first coordinate of the mixture mean remained separated across retained draws after applying the ordering rule and by visual inspection of trace plots for the key estimands.
\noindent The $\hat{R}$ values are close to 1 and the effective sample sizes are adequate for posterior summarization of the reported estimands. Below, we plot the four MCMC chains for the estimands estimated in the CMAI and antipsychotic-use analyses.

Latent implementation strata diagnostics are shown in Table~\ref{tab:latent_class_diagnostics}.

\begin{table}[H]
{\centering
\renewcommand{\arraystretch}{0.85}
\setlength{\tabcolsep}{4pt}
\begin{tabular}{lcccc}
\hline
Outcome & Stratum & \shortstack{Class size mean\\(95\% CrI)} & \shortstack{Ambiguous\\facilities$^\dagger$} & \shortstack{Relabeling gap mean\\(95\% CrI)}\tabularnewline
\hline
CMAI & $S=1$ & 33.7 [31, 37] & 3 (5.6\%) & 2.15 [1.46, 2.84]\tabularnewline
& $S=2$ & 20.3 [17, 23] & 3 (5.6\%) & 2.15 [1.46, 2.84]\tabularnewline
\hline
Antipsychotics & $S=1$ & 34.3 [30, 39] & 5 (9.3\%) & 2.15 [1.43, 2.87]\tabularnewline
& $S=2$ & 19.7 [15, 24] & 5 (9.3\%) & 2.15 [1.43, 2.87]\tabularnewline
\hline
\end{tabular}
\begin{minipage}{13cm}\linespread{1.0}
\footnotesize
$^\dagger$ A facility is classified as ambiguous if its maximum posterior class-membership probability is less than 0.70. The relabeling gap is the posterior separation between the first coordinate of the ordered mixture means; positive values away from zero indicate stable ordering of the two latent implementation strata.
\end{minipage}
\par}
\linespread{1.0}
\belowtablecaption{Posterior diagnostics for the two latent implementation strata in the METRIcAL outcome analyses. Class sizes summarize the posterior number of facilities assigned to each latent stratum after relabeling. Ambiguous facilities summarize posterior uncertainty in facility-level stratum assignment. The relabeling gap summarizes the separation in the mixture-coordinate used to order the latent strata across retained posterior draws.}
\label{tab:latent_class_diagnostics}
\end{table}

\subsection*{Sensitivity to the number of latent implementation strata}

As a sensitivity analysis, we also fit reduced models with $K=1$, corresponding to a specification without latent implementation strata. Table~\ref{tab:k_sensitivity} compares the marginal ITT and CACE estimates from the reduced $K=1$ models with the primary $K=2$ analyses.

\begin{table}[H]
{\centering
{
\renewcommand{\arraystretch}{0.85}
\setlength{\tabcolsep}{5pt}
\begin{tabular}{llccc}
\hline
Outcome & Model & Estimand & Mean (SD) & 95\% CrI\tabularnewline
\hline
CMAI & $K=1$ & ITT & 1.10 (0.95) & [-0.80, 2.95]\tabularnewline
& $K=2$ & ITT & 1.13 (0.85) & [-0.55, 2.86]\tabularnewline
& $K=1$ & CACE & 2.94 (2.51) & [-2.20, 7.74]\tabularnewline
& $K=2$ & CACE & 2.98 (2.21) & [-1.49, 7.29]\tabularnewline
\hline
Antipsychotics & $K=1$ & ITT & -2.7 (1.6) & [-6.0, 0.5]\tabularnewline
& $K=2$ & ITT & -2.9 (1.6) & [-6.1, 0.3]\tabularnewline
& $K=1$ & CACE & -7.2 (4.5) & [-16.4, 1.4]\tabularnewline
& $K=2$ & CACE & -7.8 (4.4) & [-16.9, 0.7]\tabularnewline
\hline
\end{tabular}
}
\par}
\linespread{1.0}
\belowtablecaption{Sensitivity analysis comparing marginal super-population ITT and CACE estimates from reduced $K=1$ models and the primary $K=2$ latent implementation strata models. The $K=1$ model removes latent implementation strata and is used as a reduced-model sensitivity analysis rather than as a replacement for the primary $K=2$ analysis. Antipsychotic-use estimates are reported on the percentage-point scale.}
\label{tab:k_sensitivity}
\end{table}

The marginal ITT and CACE estimates are similar under $K=1$ and $K=2$ for both outcomes. This suggests that the marginal conclusions are not driven by the inclusion of latent implementation strata, although the $K=2$ model remains the primary analysis because it also provides inference on heterogeneity across latent implementation strata. We also fit a model with $K=3$, but one latent implementation stratum remained empty in the fitted analysis. We therefore do not report $K=3$ stratum-specific causal estimates, because the empty stratum provides no empirical information for estimating its effects.

Additional posterior predictive checks are shown in Table~\ref{tab:expanded_ppc}. These checks summarize model fit for compliance rates, outcome distributions, latent class sizes, and the cluster-level variables used in the latent implementation model.

\begin{table}[h]
{\centering
{
\small
\renewcommand{\arraystretch}{0.85}
\setlength{\tabcolsep}{3pt}
\begin{tabular}{llllccc}
\hline
Outcome & Component & Discrepancy & Stratum & Study & Replicated mean & PPPV\tabularnewline
\hline
\multicolumn{7}{l}{CMAI}\tabularnewline
& Compliance & Compliance rate & All & 0.371 & 0.378 & 0.548\tabularnewline
& Outcome & Mean & All & -1.128 & -1.031 & 0.534\tabularnewline
& Outcome & Variance & All & 305.498 & 313.141 & 0.642\tabularnewline
& Compliance & Compliance rate & $S=1$ & 0.323 & 0.336 & 0.590\tabularnewline
& Outcome & Mean & $S=1$ & -2.347 & -2.355 & 0.498\tabularnewline
& Outcome & Variance & $S=1$ & 281.633 & 310.098 & 0.875\tabularnewline
& Class size & Facilities & $S=1$ & 33.701 & 32.890 & 0.406\tabularnewline
& Compliance & Compliance rate & $S=2$ & 0.448 & 0.444 & 0.478\tabularnewline
& Outcome & Mean & $S=2$ & 0.809 & 1.028 & 0.562\tabularnewline
& Outcome & Variance & $S=2$ & 337.027 & 308.962 & 0.188\tabularnewline
& Class size & Facilities & $S=2$ & 20.299 & 21.110 & 0.514\tabularnewline
& Implementation & Treated-facility $C$ mean & $S=1$ & 0.695 & 0.700 & 0.516\tabularnewline
& Facility & $Z$ mean & $S=1$ & -19.148 & -19.146 & 0.492\tabularnewline
& Implementation & Treated-facility $C$ mean & $S=2$ & -1.388 & -1.437 & 0.457\tabularnewline
& Facility & $Z$ mean & $S=2$ & 31.806 & 31.770 & 0.500\tabularnewline
\hline
\multicolumn{7}{l}{Antipsychotics}\tabularnewline
& Compliance & Compliance rate & All & 0.368 & 0.374 & 0.540\tabularnewline
& Outcome & Mean & All & 0.282 & 0.284 & 0.537\tabularnewline
& Outcome & Variance & All & 0.203 & 0.203 & 0.537\tabularnewline
& Compliance & Compliance rate & $S=1$ & 0.317 & 0.330 & 0.585\tabularnewline
& Outcome & Mean & $S=1$ & 0.257 & 0.273 & 0.752\tabularnewline
& Outcome & Variance & $S=1$ & 0.191 & 0.198 & 0.752\tabularnewline
& Class size & Facilities & $S=1$ & 34.312 & 33.534 & 0.402\tabularnewline
& Compliance & Compliance rate & $S=2$ & 0.452 & 0.447 & 0.465\tabularnewline
& Outcome & Mean & $S=2$ & 0.322 & 0.300 & 0.263\tabularnewline
& Outcome & Variance & $S=2$ & 0.219 & 0.209 & 0.264\tabularnewline
& Class size & Facilities & $S=2$ & 19.688 & 20.466 & 0.512\tabularnewline
& Implementation & Treated-facility $C$ mean & $S=1$ & 0.694 & 0.706 & 0.520\tabularnewline
& Facility & $Z$ mean & $S=1$ & -18.010 & -18.007 & 0.497\tabularnewline
& Implementation & Treated-facility $C$ mean & $S=2$ & -1.397 & -1.437 & 0.463\tabularnewline
& Facility & $Z$ mean & $S=2$ & 31.481 & 31.691 & 0.514\tabularnewline
\hline
\end{tabular}
}
\par}
\linespread{1.0}
\belowtablecaption{Expanded posterior predictive checks for the METRIcAL outcome analyses. The Study and Replicated mean columns report discrepancies averaged across posterior draws for the study and replicated datasets, respectively. Study discrepancies use draw-specific latent strata and imputed compliance or outcomes where needed. The posterior predictive $p$-value (PPPV) is the proportion of draws with both discrepancies available for which the replicated discrepancy exceeds the study discrepancy. The logit-transformed clinical proportion ($C$) is summarized among treated facilities. Average daily census ($Z$) is centered across facilities.}
\label{tab:expanded_ppc}
\end{table}

\begin{figure}[p]
 \centering
\includegraphics[width=0.70\linewidth]{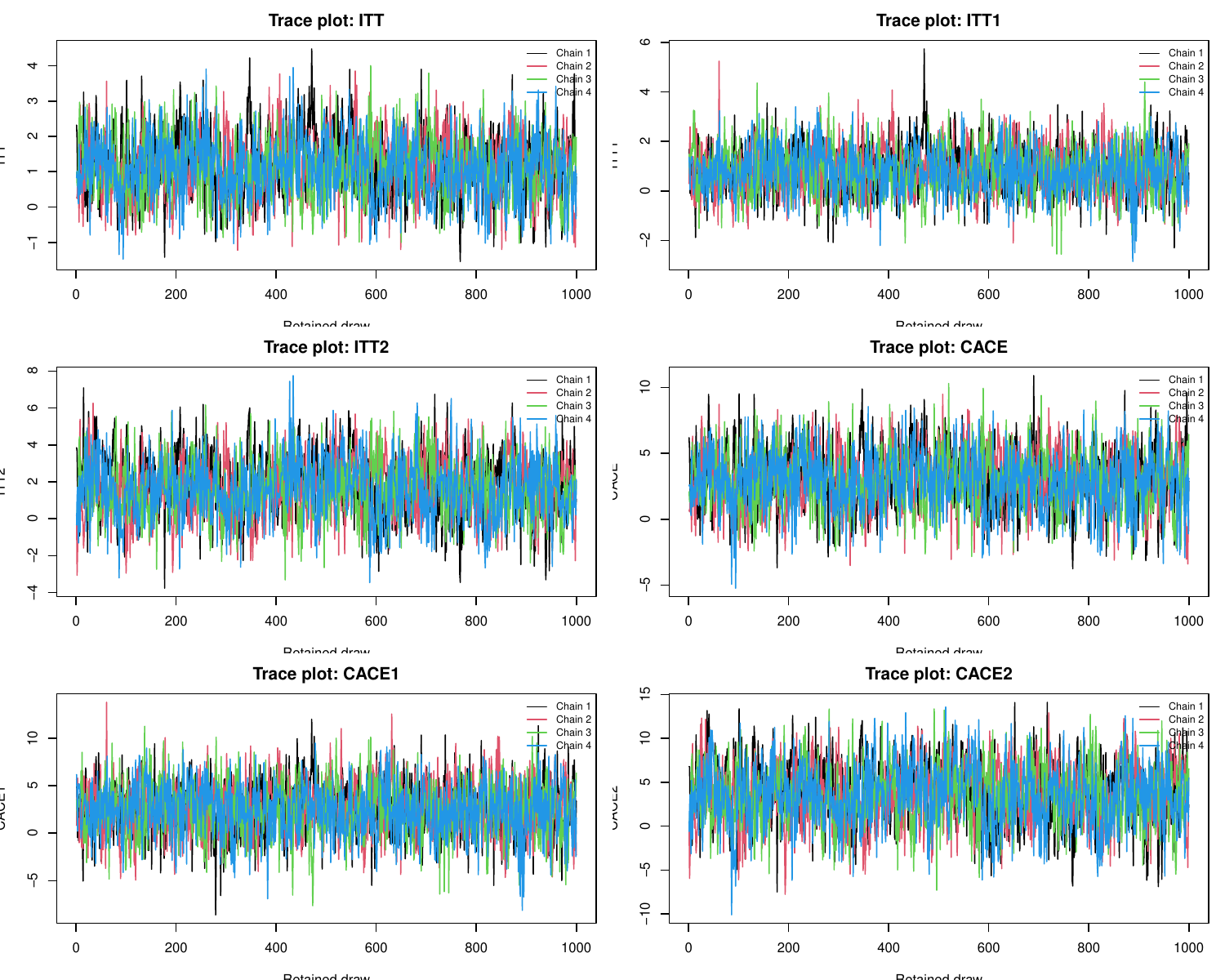}
\linespread{1.0}
     \caption{Overlaid trace plots of the 4 MCMC chains for the retained posterior samples used for inference on the CMAI outcome in Section~\ref{CMAIOutcome}.}
    \label{fig:cmai_trace_plots}

\medskip
\includegraphics[width=0.70\linewidth]{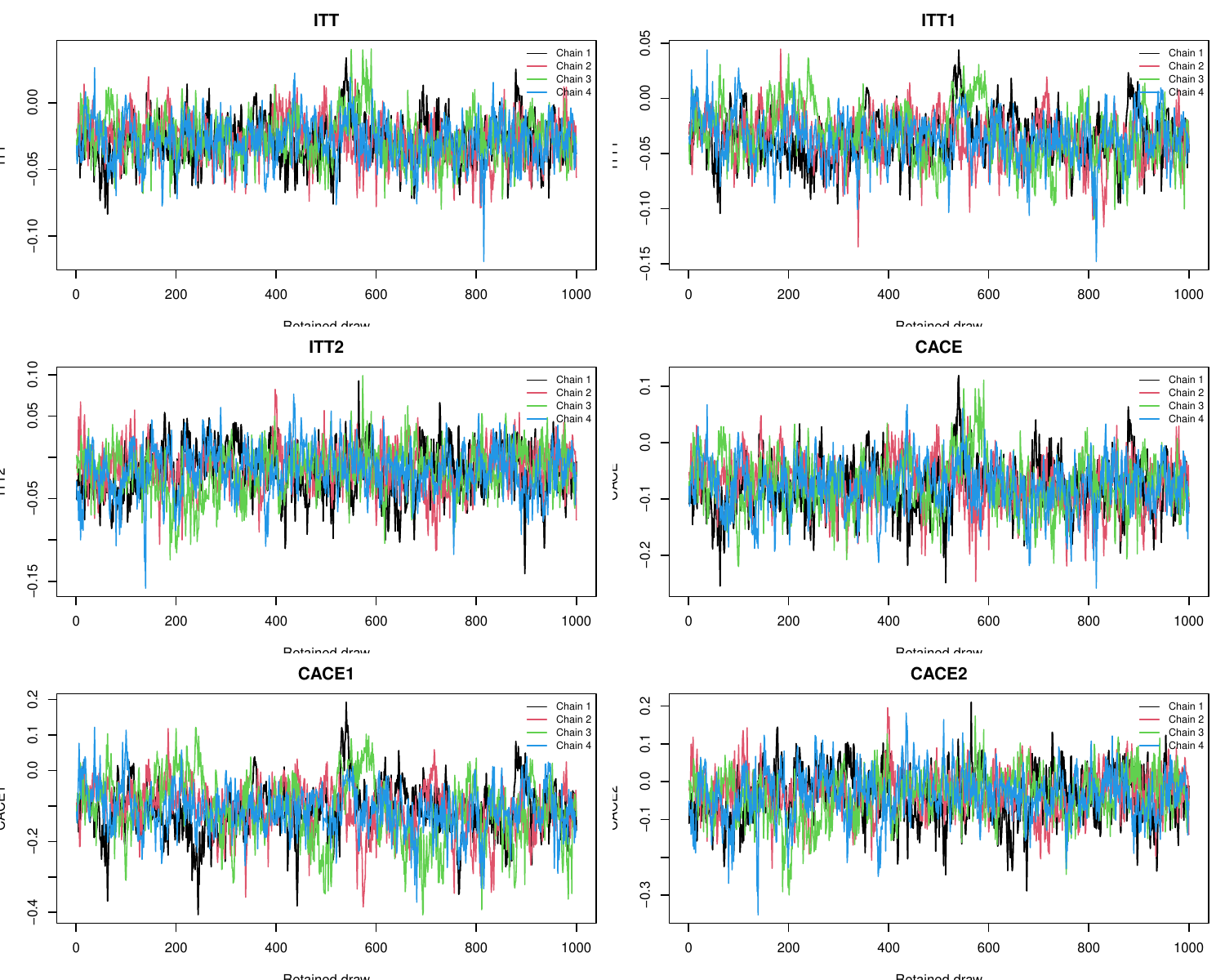}
     \caption{Overlaid trace plots of the 4 MCMC chains for the retained posterior samples used for inference on the antipsychotic-use outcome in Section~\ref{DataAnalysis}.}
    \label{fig:antipsychotic_trace_plots}
\end{figure}

\end{document}